\documentclass[sn-apa]{sn-jnl}

\usepackage{graphicx}%
\usepackage{multirow}%
\usepackage{amsmath,amssymb,amsfonts}%
\usepackage{amsthm}%
\usepackage{mathrsfs}%
\usepackage[title]{appendix}%
\usepackage{xcolor}%
\usepackage{textcomp}%
\usepackage{manyfoot}%
\usepackage{booktabs}%
\usepackage{algorithm}%
\usepackage{algorithmicx}%
\usepackage{algpseudocode}%
\usepackage{listings}%
\newcommand{\Hcal}{\mathcal{H}}
\newcommand{\Mcal}{\mathcal{M}}
\newcommand{\Ocal}{\mathcal{O}}
\newcommand{\Tcal}{\mathcal{T}}
\newcommand{\Scal}{\mathcal{S}}
\newcommand{\Vcal}{\mathcal{V}}
\newcommand{\Ucal}{\mathcal{U}}

\newcommand{\R}{\mathbb{R}}
\newcommand{\E}{\mathbb{E}}
\newcommand{\V}{\mathbb{V}}
\newcommand{\muref}{\mu^{\text{ref}}}
\newcommand{\mubar}{\bar{\mu}}
\newcommand{\Hbar}{\bar{H}}
\definecolor{darkgreen}{rgb}{0.1,.8,0.1}
\definecolor{darkred}{rgb}{0.8,.1,0.1}
\definecolor{darkblue}{rgb}{0.1,.1,0.8}

\newtheorem{theorem}{Theorem}
\newtheorem{proposition}[theorem]{Proposition}%
\newtheorem{lemma}[theorem]{Lemma}%
\newtheorem{remark}{Remark}%

\newtheorem{definition}{Definition}%

\begin{document}

\title[High order Universal Portfolio]{High order universal portfolios}

\author{\fnm{Gabriel} \sur{Turinici}}\email{Gabriel.Turinici@dauphine.fr}

\affil{\orgdiv{CEREMADE}, \orgname{Universit\'e Paris Dauphine - PSL} \\ \orgaddress{\street{Place du Marechal de Lattre de Tassgny}, \city{Paris}, \postcode{75116}, \state{Paris}, \country{FRANCE}} \\ \ \\
Initial version: November 2023; this version: May 27th, 2025
}

\abstract{The Cover universal portfolio 
has many interesting theoretical and numerical properties and was investigated for a long time. Building on it, we explore what happens when we add this UP to the market as a new synthetic asset and construct by recurrence higher order UPs. 
We investigate some important theoretical properties of the high order UPs and show in particular that they are indeed different from the Cover UP and are capable to break the time permutation invariance. 
We show that under some perturbation regime the second  order UP has better Sharp ratio than the standard UP and briefly investigate arbitrage opportunities thus created. Numerical experiences on a benchmark from the literature confirm that high order UPs improve Cover's UP performances.}




\maketitle

\section{Motivation and literature review}\label{sec:intro}

The introduction of the Universal Portfolio (abbreviated ``UP'' from now on) 
algorithm in \cite{cover91} generated a considerable amount of interest in the quantitative portfolio management community because of its theoretical and practical appeals: the universal portfolio 
can be constructed {\bf without any
	 information on the future price evolution nor any statistical assumption }
but is proven to reach asymptotically a performance compatible with the best constant rebalanced portfolio (denoted by ``BCRP'' in the sequel) {\bf chosen with the benefit of hindsight} \cite[Thm. 6.1 p.13]{cover91}~\footnote{In more precise terms it was proved under some technical assumptions that $\lim_{T\to \infty}\frac{\ln(UP^1_T)-\ln(CRP(w^\star)_T)}{T}=0$, see Remark~\ref{rem:optimality} for notations.} and in particular is asymptotically not worse 
than any individual asset. While the BCRP cannot be implemented because requires future data, the UP is implementable at each time step.
Tested on some simple benchmarks this strategy proved to be efficient if given enough time to reach the asymptotic regime. 

Several works explored various aspects of the theory: 
 \cite{cont_time_univ_portf_jamshidian92} 
wrote a continuous time version  and gave further results under log-normal assumptions, 
\cite{Helmbold98} proposed a online reinforcement learning-style  version of the algorithm, 
\cite{li_moving_avg_portf_2015} continued in this direction under the assumption of moving average reversion while \cite{blum1997universal} investigated the theoretical and practical impact of transaction costs. 
A connection with  the general framework of stochastic portfolio theory see  \cite{fernholz2002stochastic} was proposed by  \cite{stoch_portf_up_num18} together with a comparison with the the ``num\'eraire'' portfolio.
A more general view on the learning rate is presented by \cite{vovk1998universal}.
\cite{cover1996universal}
 came back to the subject opening the discussion of how to incorporate additional side information and later \cite{cover98} explored derivative pricing related to the UP.

On the computational side, \cite{kalai2002efficient} showed that an astute sampling of the set of Constant Rebalanced Portfolios (``CRP'' from now on) can give rise to efficient algorithms for universal portfolios; and the work continues 
 up to this day with \cite{parthasarathy2023online} investigating the optimal way to assign weights to different CRPs. 
  For further results see also the review of \cite{li2014online_survey}.

It is thus clear that the universal portfolio can provide interesting insights into investing strategies; given its importance, it is then natural to see the universal portfolio as a kind of synthetic asset (to be added to the same market as the primary assets used to build it) in the same vein as market indices and associated ETFs are nowadays an important instruments for market gauging.
For a general set of market assets $\Mcal$ denote by $UP(\Mcal)$, denoted in the following
$UP^1(\Mcal)$, the universal portfolio associated to $\Mcal$\footnote{When there is no ambiguity we will also denote it by $UP(\Mcal)$.}. 
A natural question is what happens if we add $UP(\Mcal)$ to the market $\Mcal$. 
One can define recursively the universal portfolio
$UP^2(\Mcal)$ 
as the 
universal portfolio of the market $\Mcal$ to which we add the synthetic asset $UP(\Mcal)$~: 
$UP^2(\Mcal):= UP(\Mcal \cup \{UP(\Mcal) \})$ ; this procedure can be continued recursively~:  
$UP^3:= UP(\Mcal \cup \{UP(\Mcal)\} \cup \{UP^2(\Mcal)\})$ 
 and so on. 
 As a matter of vocabulary, we will call such portfolios ``high order universal portfolios'', abbreviated HOUP.
 Several questions are now in order : 
\begin{itemize}
	\item do $UP^2(\Mcal)$, $UP^3(\Mcal)$ bring anything new with respect to $UP^1(\Mcal)$ i.e., are they different from $UP^1(\Mcal)$?
	\item if we iterate this construction $\ell$ times resulting in the $\ell$-th order $UP^\ell(\Mcal)$ portfolio, how does its performance compares with that of $UP^1(\Mcal)$ ?
\end{itemize}
The purpose of this paper is to answer such questions. The outline of the work is the following~: in section~\ref{sec:notations} we introduce formally the high order universal portfolios followed in section~\ref{sec:theory} 
by some theoretical results
showing that the HOUP indeed differs from standard UP. We continue in section \ref{sec:theortical_cont_time}  with additional perturbative continuous time results 
about the Sharpe ratio of the HOUP compared with plain UP. In section~\ref{sec:numerical} we present the performance of the HOUPs on several benchmarks from the literature followed in section~\ref{sec:conclusion} by concluding remarks.

\section{High order universal portfolios (HOUP): definition and notations}\label{sec:notations}

Consider $\Mcal$ a set of $K$ financial assets and a time frame of $N$ instants $t\in \Tcal :=\{1,...,T\}$;~\footnote{Of course, everything said here can be extended to a general time frame
$\{t_1,...,t_N=T\}$ at the price of writing $t_i$ instead of $i$ and $t_{i+1}$ instead of $t+1$... } denote by $S^k_t$ the price of the $k$-th asset at time $t$; by convention $S^k_0=1$ for all $k\le K$.
Introduce the price relatives
$f^k_t = S^{k}_{t} /  S^{k}_{t-1}$ for any $t\in \Tcal$.\footnote{It is assumed that all asset prices are strictly positive at any time.} In financial jargon quantities $r^k_t=f^k_t-1$ are also called ``returns'' of the assets.

A constant rebalanced portfolio (CRP) is defined by a set of $K$ non-negative  weights $w_1,..., w_K$ that sum up to one. For each such $w$ the portfolio $CRP(w)$ is such that at any time $t$ the quotient of the wealth invested in the asset $k$ with respect to total portfolio value is $w_k$. In financial literature CRP is also called a ``Constant Mix'' portfolio.

Note that CRP is a {\bf dynamic} investment strategy because at any time $t\ge 0$ one has to {\bf rebalance} the allocation that may have drifted because of the change in prices: imagine for instance a market with $K=2$ assets and a $CRP(1/2,1/2)$ allocation at $t=0$. If at $t=1$ one of the assets rises significantly and the other decreases  significantly, the amount of wealth invested in the first asset will become much larger than that invested in the second asset. In this case a rebalancing is done at the prices available at time $t=1$ to restore the proper allocation proportions $(1/2,1/2)$. Note in particular that CRP is distinct from the so-called ``Buy\&Hold'' (also known as ``Split-and-Forget'') portfolio that invests at $t=0$ equal parts of the initial wealth in each asset and do not trade at all afterwards (no rebalancing).

Supposing a portfolio starting from value $1$ at time $0$ and 
denoting $f_t$ the vector with components $f^k_t$, $k\le K$, 
the value $\Vcal(w)_t $ at time $t \in \Tcal$ of the $CRP(w)$ portfolio will be~:
\begin{equation}
\Vcal(w)_0=1, \  \forall t \in \Tcal~:~ \Vcal(w)_t = \prod_{s=1}^{t} \langle w, f_s\rangle.
\label{eq:definitionVw}
\end{equation}
In ``returns'' formulation this reads
$\Vcal(w)_t = \prod_{s=1}^{t} (1+\langle w, r_s\rangle)$. 
Denote $\Scal_K$ the unit simplex of dimension $K$~:
\begin{equation}
\Scal_K = \{w = (w_k)_{k=1}^K \in \R^K : w_k \ge0, \sum_{k=1}^K w_k = 1\}.
\label{eq:simplex}
\end{equation}
We will introduce a distribution on $\Scal_K$; in the initial proposal Cover used the uniform distribution but in general Dirichlet laws have also been advocated. All that is said in the following extends to any of these distributions but for simplicity we will use the uniform law over $\Scal_K$ and denote $\E_w[\cdot]$ or $\E_{w \in \Scal_K}[\cdot]$ the average over it.

\begin{definition}[Cover universal portfolio]
With the notation \eqref{eq:definitionVw} 
the universal portfolio for the market $\Mcal=\{ S^k; k \le K \}$ is defined to have the  allocation $u_t \in \Scal_K$ at time $t$:
\begin{equation}
	u_0=(1/K,...,1/K), \ u_t = \frac{\E_w[ w \cdot \Vcal(w)_t]}{\E_w [\Vcal(w)_{t}]} \textrm{ for } t 
	\ge 0,
	\label{eq:defup}
\end{equation}
i.e. the allocation is a weighted average of $w\in \Scal_K$ with weights proportional to the  performances $\Vcal(w)_t$ up to time $t$.
\end{definition}

For convenience, we denote $UP^1$ the universal portfolio thus defined and in particular $UP^1_t$ will be its value at time $t$. Note that the construction of $UP^1$ does not requires any forward information on the market $\Mcal$ and can be realized on the fly as time advances. 
In particular one uses $f_1$ to compute $u_1$ and $f_2$ to  compute $u_2$ and so on. The gains of the portfolio itself result from data $f_1$ applied to allocation $u_0$ and data $f_2$ applied to allocation $u_1$ and so on.
\begin{definition}[high order universal portfolios - HOUP]
	Let $\Mcal$ be a market and denote $UP(\Mcal)$ its universal portfolio. Then order $\ell \ge 1$ universal portfolios of the market $\Mcal$ are denoted $UP^{\ell}$ and defined recursively by~: 
\begin{equation}
	UP^{\ell}= UP(\Mcal^\ell), \textrm{ where } 
	\Mcal^1 = \Mcal, \ 
	\Mcal^{\ell}= \Mcal^{\ell-1} \cup \{UP^{\ell-1}\} \textrm{ for } \ell \ge 2.
\end{equation} 
\end{definition}
\begin{remark}
	It can be proved from \eqref{eq:up_as_average} and was documented in in \cite{cover91} that the value $UP^1_t$ of the universal portfolio is the average of the values of all possible CRPs~:
	\begin{equation}
		UP^1_t= \E_w \Vcal(w)_t. \label{eq:up_as_average}
	\end{equation}
	This property is the basis of its asymptotic performance because any  $CRP(w^\star)$ 
	that is optimal in hindsight (i.e., $CRP(w^\star)$=BCRP) corresponds to some $\Vcal(w)$ with $w\in \Scal_K$ and 
	will end up imposing its growth rate to all other members of the average above.
 In fact as put by \cite{li_moving_avg_portf_2015} the UP operates as a Fund of Funds (FoF), each elementary fund corresponding to a $CRP(w)$ strategy. In such a view, the $CRP(w^\star$) is already a member of the UP portfolio, so the FOF will benefit from its gains.  
	 Since we are enlarging the market at any step, any $UP^\ell$ will keep the same property of optimality  (this can be formalized mathematically). In particular, since $UP^1$ is in the market used to obtain $UP^2$ we expect that the performance of $UP^2$ is at least as good as that of $UP^1$ and in general we expect the performance to improve when $\ell$ increases. This theoretical point will be checked empirically in section~\ref{sec:numerical}.
\label{rem:optimality}
\end{remark}

\begin{remark}
	To implement $UP^\ell$ for $\ell>1$ one does not need anything more than the access to the market $\Mcal$ as was the case for $UP$.
Adding UP to the initial market $\Mcal$ is a ``thought experiment'', in practice
each $UP^\ell$ can be expressed as a portfolio containing only assets of the initial market $\Mcal$; note however that {\bf in general $UP^\ell$ is not a CRP} because $u_t$ from equation~\eqref{eq:defup} is not necessarily constant in time (neither for UP nor for $UP^\ell$, $\ell >1$). In particular, while equation \eqref{eq:up_as_average} shows that for   $\ell=1$ $UP^\ell \le CRP(w^\star)$,  when $\ell >1$ \eqref{eq:up_as_average} involves an average over $S_{K+\ell-1}$ and does allow to conclude that $UP^\ell$ is  necessarily bounded by $CRP(w^\star)$ ($w^\star \in \Scal_K$).
\label{rem:other_measures}
\end{remark}

\begin{remark}[other measures over the simplex]
Note that formula \eqref{eq:up_as_average} can also be used as definition of the universal portfolio when the measure over the unit simplex $\Scal_K$ is not uniform. 
For instance \cite{cover1996universal} use a $Dirichlet(1/2, ..., 1/2)$ measure. Each measure will give another ``flavor'' of Universal portfolio. 
Let us take the two extreme ones: the $Dirichlet(0, ..., 0)$ distribution is only supported in the vertices of the simplex so in this case the UP is simply the ``Buy\&Hold'' portfolio.
On the contrary, when $\alpha \to \infty$ the distribution  $Dirichlet(\alpha, ..., \alpha)$ is increasingly concentrated at the center of the simplex. In this case, the same formula says that in the limit the Universal portfolio is the uniform $CRP(1/K, ..., 1/K)$.
\end{remark}

\subsection{High order portfolios for other online strategies}
\label{sec:general_high_order_online_strat}

The definition of the HOUP leads to the following general question: take some online strategy $\Ocal$ that depends on the assets of the market $\Mcal$ and iterate the high order procedure described above by considering a sequential set of virtually enlarged  markets starting from 
from  $\Mcal_\Ocal^1= \Mcal$ and constructed by recurrence 
by adding $ \Ocal(\Mcal^{\ell-1}_\Ocal)$ to $ \Mcal^{\ell-1}_\Ocal$ :
 $\Mcal_\Ocal^\ell= \Mcal^{\ell-1}_\Ocal \cup \{  \Ocal(\Mcal^{\ell-1}_\Ocal) \}$ for $\ell \ge 2$. The question is whether  $\Mcal_\Ocal^\ell$ is at all different from $\Mcal$ and if it is useful from the investor's point of view.

The answer to the first question is surprising in the remarkable cases of
 ``Buy\&Hold'' and uniform CRP portfolios~: 
 take first the 
 ``Buy\&Hold'' strategy denoted $\Ocal = BH$; then 
 $\Mcal_{BH}^\ell=\Mcal^{\ell-1}_{BH}$ as soon as $\ell \ge 3$. Indeed, 
  the second order
 ``Buy\&Hold'' portfolio  $BH(\Mcal^2_{BH})$, that operates on $\Mcal^2_{BH}$ will divide initial investment equally between all assets of the market $\Mcal^1_{BH}=\Mcal$ and the (first order)
 ``Buy\&Hold'' portfolio $BH(\Mcal)$; but $BH(\Mcal)$ is already an uniform investment over the assets in $\Mcal$ so finally  $BH(\Mcal^2_{BH})$ remains a uniform investment over the assets in $\Mcal$ which means that  $BH(\Mcal^2_{BH})= BH(\Mcal)$ and the recurrence stops here.
  A similar situation arrives when we consider $\Ocal$ to be the uniform CRP portfolio. In this case the allocation changes with time but a similar reasoning shows that at each time step the return of the second order uniform CRP remains the average of the returns of the assets in the market $\Mcal$ as was the case for the first order CRP. So with the previous notations 
  we have  $CRP(\Mcal^2_{CRP})= CRP(\Mcal)$ and again the recurrence stops at $\ell=2$.
 
Other online strategies may have though a different fate. For instance this invariance also fails 
for the ``follow the leader'' (FTL) approach
or the ``exponential gradient'' (EG) strategy\footnote{See \cite{li2014online_survey} for a mathematical presentation of all these online allocation protocols.}, although rigorous proofs are not available at this time. 

\section{Theoretical questions}\label{sec:theory}

\subsection{HOUP versus UP}

Besides the optimality discussed in Remark~\ref{rem:optimality}  
an important preliminary question is whether $UP^2$ is different at all  from $UP^1$, i.e., does the introduction of the synthetic asset $UP^1$ brings any new information and changes the 
way to compute the universal portfolio. 
Note that intuitively, since universal portfolios are obtained by averaging existing assets,
one could expect that introducing the average into the market would not change anything at all. The somewhat counter-intuitive result is that, because here averages are non-linear functions of the returns\footnote{In the construction of the UP, the returns up to time $t$ enter linearly in the value when $t=1$ but enter non-linearly at times $t\ge2$; compare for instance formulas \eqref{eq:up1_1} and \eqref{eq:up1_2}.}
 computing high order universal portfolios  enrich the set of possible strategies. The formal answer is given in the following proposition~: 

\begin{proposition}
For a general market $\Mcal$ the higher order universal portfolios $UP^\ell$, $\ell \ge1$ are not all equal to $UP^1$.
\label{prop:houp_not_all_equal}
\end{proposition}

\begin{remark}(other measures on the simplex: follow-up) 
If we recall the remark	\ref{rem:other_measures} we understand that this assertion is not really trivial.
We remind that $\alpha$ denotes the (constant) parameter of the Dirichlet distribution. 
When $\alpha=0$ all $UP^\ell$, $\ell \ge 1$ will all be equal to the ``Buy\&Hold'' portfolio. When $\alpha\to \infty$ all  $UP^\ell$ will be equal to the uniform $CRP$. So, while both extreme cases are degenerate we prove that for $\alpha=1$ (that is the uniform distribution) the procedure does create new portfolios.
\end{remark}

\begin{remark} 
The formulation of the result appears somehow awkward, let us explain why it has its present form. Take the statement :  $UP^\ell  \neq UP^{\ell'}$. For given horizon $T$, the equalities 
$UP^\ell_t = UP^{\ell'}_t$ for {\bf all} $t=1,...,T$ form
a system of $T$ nonlinear equations with $T \cdot K$ unknowns $f^k_t$. Since the system is under-specified (i.e., there are more unknowns than equations), 
it is very likely that 
for given, fixed, $\ell$ and $\ell'$
there exists at least one market $\Mcal$ where  we have
{\bf $UP^\ell = UP^{\ell'}$ for all times $t$ up to time $T$} (and even for $T=\infty$). 
It may even be possible that $UP^\ell = UP^{\ell'}$ for all $t\le T$ and all $\ell, \ell' \le K$ !  So some care is needed when dealing with this statement.
 \label{rem:continuity}
\end{remark}

\begin{proof}
We will proceed by contradiction. Suppose on the contrary that for all possible markets 
$\Mcal$ 
and all $\ell \ge 1$, all
$UP^\ell$ are equal to $UP^1$. In particular consider the simple situation when $K=2$. 
To ease notations we will denote $a_t = f^1_t$, $b_t = f^2_t$ for all $t$.
In this case the expectation with respect to the uniform law over the simplex $\Scal_2$ can be computed as an integral over $x\in [0,1]$ taking $w=(x,1-x)\in \Scal_2$.
According to its definition, the allocation of $UP^1$ is $(1/2,1/2)$ at time $0$; at time $t=1$ its allocation will change to~:
\begin{equation}
u_1^1 = \frac{\E_w[ w \cdot \Vcal(w)_t]}{\E_w [\Vcal(w)_1]} = 
\frac{\int_0^1 (x,1-x) \cdot (x a_1 + (1-x)b_1) dx}{
\int_0^1 (x a_1 + (1-x)b_1) dx}= \frac{(2a_1+b_1,2b_1+a_1)}{6(a_1+b_1)}.
\end{equation} 
Its value will be 
\begin{align}
UP^1_1 =  \frac{a_1+b_1}{2}, \label{eq:up1_1}
\end{align}
which is the same as its first price relative that we will denote $c_1$~: $c_1:=(a_1+b_1)/2$. We will denote in general by $c_t$ the price relative of the first universal portfolio $UP^1$. The value at time $t=2$ of $UP^1$ is then~:
\begin{eqnarray}
& \ & 
UP^1_2 = \E_w \Vcal(w)_2 =\int_0^1 (x a_1+(1-x)b_1) (x a_2+(1-x)b_2) dx
\nonumber \\ & \ & 
=\frac{2 a_1a_2+ 2b_1 b_2+ a_1b_2+ a_2 b_1}{6}.   \label{eq:up1_2}
\end{eqnarray}
This informs that $c_2$, the price relative of the portfolio $UP^1$ from time $1$ to time $2$ is
\begin{equation}
c_2=  \frac{2 a_1a_2+ 2b_1 b_2+ a_1b_2+ a_2 b_1}{6} \cdot \frac{1}{c_1}=   \frac{2 a_1a_2+ 2b_1 b_2+ a_1b_2+ a_2 b_1}{3(a_1+b_1)}.
\label{eq:c2}
\end{equation}
 For convenience the price relative of the market   $\Mcal^2$ are recalled in the table~\ref{table:pricerelativest2}.

\begin{table}[h]
\caption{Price relatives of the market $\Mcal^2$.}\label{table:pricerelativest2}%
\begin{tabular}{@{}l|lll@{}}
\toprule
time transition& Asset 1 & Asset 2  & Asset 3 = $UP^1$ \\
\midrule
$t=0\to1$ & $a_1$  & $b_1$   & $c_1=\frac{a_1+b_1}{2}$ \\
$t=1\to2$ & $a_2$    & $b_2$   & $c_2=\frac{2 a_1a_2+ 2b_1 b_2+ a_1b_2+ a_2 b_1}{3(a_1+b_1)}$ \\
\botrule
\end{tabular}
\end{table}

On the other hand, when $UP^1$ is added to the market, the value of $UP^2$ at time
$t=0$ will be $UP^2_0=1$ (starting value), at time $t=1$ will be $(a_1+b_1)/2$ and at time 
$t=2$ will be~: 
\begin{eqnarray}
	& \ & 
	UP^2_2 = \E_{(w_1,w_2,w_3)\in \Scal_3}[ (w_1 a_1+w_2 b_1+ w_3 c_1) (w_1 a_2+w_2b_2+ w_3 c_2)]
	\nonumber \\ & \ & 
	= 	\frac{2 a_1a_2 + 2 b_1 b_2 + 2 c_1 c_2+ a_1 b_2 + a_2b_1+ a_1c_2 +a_2 c_1 + b_1 c_2 + b_2 c_1}{12}
	\nonumber \\ & \ & 
= 	\frac{23 a_1a_2 + 23 b_1 b_2 + 13a_1 b_2 + 13a_2b_1}{72}, \label{eq:up2_2}
\end{eqnarray}
where we used relation $c_1=\frac{a_1+b_2}{2}$, equation~\eqref{eq:c2} and 
 identities 
\begin{equation}
\E_{(w_1,w_2,w_3)\in \Scal_3} [w_i w_j] = 1/12 + 1/12 \delta_{ij}, \forall i,j \in \{1,2,3\}.
\end{equation} 
The computation of the last identity was checked in full detail by hand by the author. However intermediary steps 
are cumbersome and 
will not be presented here but the reader can 
 obtain independent verification of the result using 
the small symbolic Python code listed  
in Appendix~\ref{appendix:python_code_for_proof}.
Since formulas  \eqref{eq:up1_2} and \eqref{eq:up2_2} show that in general $UP^1_2 \neq UP^2_2$ we obtain a contradiction with the assumption 
that all $UP^\ell$ equal $UP^1$, which
ends the proof.
\end{proof}

\subsection{On the permutation invariance}

To explain the next result we recall \cite{cover91} that states that the return of the UP is invariant with respect to permutations of the time instants. More precisely, consider a permutation $\sigma$ of $\Tcal$
and the market $\Mcal_\sigma$ that contains assets with prices $Y^k_t$ (we keep the convention $Y^k_0=1, \forall k\le K$) and factors 
$g^k_t = Y^{k}_{t+1}/Y^k_t$ such that $g^k_t = f^k_{\sigma(t)}$ for any $t\in \Tcal$, $k\le K$. It was proven that~: 
\begin{equation}
UP(\Mcal)_T= UP(\Mcal_\sigma)_T, 
\textrm{ for any permutation } \sigma \textrm{ of }\Tcal.
\end{equation}
Although this is an interesting property this means that UP cannot, by design, exploit time correlations or the time ordering of the factors. Some attempts to settle this problem came from the inclusion of side information, see \cite{cover1996universal}. We take here a different view and  show that higher order UP break the permutation symmetry and thus can possibly exploit the time coherence of the price relatives. 

\begin{proposition}
There exist at least one market $\Mcal$, one order $\ell >1$ and one permutation $\sigma$ of $\Tcal$ such that~: 
\begin{equation}
	UP^\ell(\Mcal)_T \neq UP^\ell(\Mcal_\sigma)_T.
\end{equation}
\label{prop:no_time_invariance}\end{proposition}
\begin{remark}Of course, the permutation invariance may still occur for some particular values of factors, see discussion in Remark~\ref{rem:continuity};
	it is even natural to expect that varying the factors $f^k_t$ one may occasionally match the values of 
	$UP^\ell(\Mcal)_T$ and $UP^\ell(\Mcal_\sigma)_T$. This explains the form of the Proposition~\ref{prop:no_time_invariance}. 
\end{remark}

\begin{proof}
We will simply present a counter-example for the case $\ell=2$ and the market in table~\ref{table:prop_time}; first two columns are given, the third is derived from them using previous formulas for $c_1$ and $c_2$. To compute $c_3$ we use that the value of $UP^1$ at time $t=3$ is given by ($x$ is uniform in $[0,1]$)~:
\begin{eqnarray}
& \ & 
UP^1_3= \E_x[(x a_1 + (1-x)b_1)(x a_2 + (1-x)b_2)(x a_3 + (1-x)b_3)]
\nonumber \\ & \ &
=\frac{(a_1+b_1)(a_2+b_2)(a_3+b_3)+ 2 a_1a_2a_3+2b_1b_2b_3}{12}. 
\end{eqnarray}

\begin{table}[h]
	\caption{Price relatives of the market $\Mcal^2$ in proposition~\eqref{prop:no_time_invariance}.}\label{table:prop_time}%
	\begin{tabular}{@{}l|lll@{}}
		\toprule
		time transition& Asset 1 & Asset 2  & Asset 3 = $UP^1$ \\
		\midrule
		$t=0\to1$ & $a_1=1$  & $b_1=2$   & $c_1=\frac{a_1+b_1}{2}=3/2$ \\
		$t=1\to2$ & $a_2=2$    & $b_2=1$   & $c_2=\frac{2 a_1a_2+ 2b_1 b_2+ a_1b_2+ a_2 b_1}{3(a_1+b_1)}=13/9$ \\
		$t=2\to3$ & $a_3=2$  & $b_1=1$   & $c_3=UP^1_3/c_2=3/2$ \\
		\botrule
	\end{tabular}
\end{table}
We now take the permutation $\sigma$ that exchanges $1$ and $3$ and consider the market $\Mcal_\sigma^2$ with entries in table~\ref{table:prop_time_sigma}. 
Note that, as expected, even if the entries $c_k$ in  table \ref{table:prop_time_sigma} are not permutation of the entries $c_k$ in table \ref{table:prop_time}, the final value of $UP^1$ is the same at time $t=3$ for both markets as proved by Cover: $(3/2)\cdot(13/9)\cdot(3/2) =13/4= (3/2)\cdot(14/9)\cdot (39/28)$.

\begin{table}[h]
	\caption{Price relatives of the market $\Mcal^2_\sigma$ in proposition~\eqref{prop:no_time_invariance}.}\label{table:prop_time_sigma}%
	\begin{tabular}{@{}l|lll@{}}
		\toprule
		time transition& Asset 1 & Asset 2  & Asset 3 = $UP^1_\sigma$ \\
		\midrule
		$t=0\to1$ & $a_1=2$  & $b_1=1$   & $c_1=\frac{a_1+b_1}{2}=3/2$ \\
		$t=1\to2$ & $a_2=2$    & $b_2=1$   & $c_2=\frac{2 a_1a_2+ 2b_1 b_2+ a_1b_2+ a_2 b_1}{3(a_1+b_1)}=14/9$ \\
		$t=2\to3$ & $a_3=1$  & $b_1=2$   & $c_3=(UP_\sigma)^1_3/c_2=39/28$ \\
		\botrule
	\end{tabular}
\end{table}
With these provisions one can compute the values of the second order universal portfolio for the two markets at time $t=3$, which will be~: 
\begin{equation}
\E_{w\in \Scal_3} [(w_1 a_1 + w_2 b_1 + w_3 c_1) (w_1 a_2 + w_2 b_2 + w_3 c_2)
(w_1 a_3 + w_2 b_3 + w_3 c_3)].
\end{equation}
To this end following formulaes are used for $i,j,k\in \{1,2,3\}$~:
\begin{equation}
\E_{w\in \Scal_3} [w_i w_j w_k] = 
\begin{cases}
\frac{1}{10} \textrm{ if } i,j,k \textrm{ are all distinct} \\
\frac{1}{60} \textrm{ if } i=j=k \\
\frac{1}{30} \textrm{ if } i=j\neq k \textrm{ and similar cases } \\
	\end{cases}.
\end{equation}
We obtain for market $\Mcal$~:  $UP^2(\Mcal)_3=3533/1080$ while for its permutation 
$\Mcal_\sigma$ we obtain $UP^2(\Mcal_\sigma)_3=49457/15120$, which are different; the proof is complete.
\end{proof}
\begin{remark}
    We do not have yet enough insight as to what properties of the patterns of the price relative are captured by the HOUP; this is nevertheless an important question as it could lead to characterize better in which  regime the HOUPs outperform ``classical'' UPs.
\end{remark}
\begin{remark}
Similar ideas can be used to extend both propositions \ref{prop:houp_not_all_equal} and \ref{prop:no_time_invariance} to other strategies including ``follow the leader'' (FTL) or ``exponential gradient'' (EG). For instance, FTL is not time invariant because the $t=1$ allocation always takes the best performing asset over the first time step so depends crucially on which value is taken first. On the other hand the fact that the addition of the first order FTL strategy to the market gives a non trivial second order  FTL is a lengthier computation in general; note that a simple numerical example will provide the conclusion, take for instance the market $\Mcal^2$ in Table~\ref{table:prop_time}. Same discussion works for EG.
\label{rem:other_strategies}
\end{remark}

{
\section{Perturbative continuous time results}\label{sec:theortical_cont_time}

We will consider now the continuous time versions of the high order universal portfolios and will work in a perturbative regime around a reference dynamics.
 More specifically, 
 assume that the market $\Mcal$ containing $K$ assets is such that~:
\begin{equation}
	\frac{d S_t^k}{S_t^k} = \mu_k^\epsilon(t) dt + d H_k^\epsilon(t),\  S_0^k=1.
\end{equation}
Here  $\epsilon$ is a perturbation parameter that
will be discussed later.
Here $\mu_k^\epsilon(t)$ are continuous deterministic functions and $H_k^\epsilon=(H_k^\epsilon(t))_{t\ge 0}$ are $L^2$ integrable martingales with finite quadratic variation starting from $0$ at time $0$. We will not consider singular settings so we suppose that regularity conditions required to write this stochastic differential equations are satisfied.
In particular the setting of  \cite[equation 2.1]{cont_time_univ_portf_jamshidian92} is realized when
$\mu_k^\epsilon(t)=\mu_k(t)$ and
$H_k^\epsilon(t) = \sum \sigma_{kj}(t) W_j$ with $\sigma(t)$ a $N\times N$  matrix, and $W_j$ independent Brownian motions.

\medskip

The $CRP(w)$ denoted, as before, by $\Vcal(w)_t$ follows the dynamics
\begin{equation}
	\frac{d \Vcal(w)_t}{\Vcal(w)_t} =\sum_{k=1}^K [w_k \mu_k^\epsilon(t) dt +  w_k  d H_k^\epsilon(t)], \ \Vcal(w)_0=1,
\end{equation} 
which leads to 
\begin{equation}
\Vcal(w)_T =exp \left\{\sum_{k=1}^K  [\int_0^T w_k \mu_k^\epsilon(t) dt +
 w_k  H_k^\epsilon(T)] - \frac{1}{2}\sum_{k,\ell\le K} w_k w_\ell \int_0^T d\langle H_k^\epsilon , H_\ell^\epsilon \rangle_t  \right\}.
\end{equation} 
We will compare below some properties of $UP^\ell(\Mcal)$ for $\ell=1,2$
 in the limit where
$\mu_k$ are close to some $\muref(t)$ and $H_k^\epsilon$ are small; more precisely, we will consider some perturbation parameter $\epsilon$ and assume~: 
\begin{equation}
\mu_k^\epsilon(t) = \muref(t) + \epsilon \mu_k(t) + o(\epsilon),
\ H_k^\epsilon = \sqrt{\epsilon} H_k + o(\epsilon).
\end{equation}
In full rigor the $ o(\epsilon)$ term above is to be understood as 
$\lim_{\epsilon \to 0} \frac{\| \mu_k^\epsilon(t) - \muref(t) - \epsilon \mu_k(t) \|_{L^2([0,T])}}{\epsilon}=0$, 
$\lim_{\epsilon \to 0} \frac{\|  H_k^\epsilon - \sqrt{\epsilon} H_k \|_{ \mathcal{L}^2([0,T])}}{\epsilon}=0$ where
$L^2([0,T])$ is the usual $L^2$ Hilbert space of functions on $[0,T]$ whose square is integrable while
$\mathcal{L}^2([0,T])$ is the corresponding space for stochastic processes (integral of the square over $[0,T]$ has finite average). This convention is used throughout  the proof below. The reader that does not want to deal with these technicalities can just consider that 
$\mu_k^\epsilon(t) = \muref(t) + \epsilon \mu_k(t)$ and $ H_k^\epsilon = \sqrt{\epsilon} H_k $.
We will prove a result stating that  $UP^2(\Mcal)$ has better Sharpe ratio than 
 $UP^1(\Mcal)$. 
We recall that for a time horizon $[0,T]$ the Sharpe ratio \cite{sharpe_ratio} of an asset with price $(A_t)_{t\ge 0}$ is
$\frac{\E \left[\ln(A(T)/A(0))\right]-rT}{std \left[\ln(A(T)/A(0))\right]}$ where $r$ is the risk-free rate.
\begin{proposition}
	Consider the market $\Mcal$ above and
	denote $\mubar= \frac{1}{K}\sum_{k=1}^K \mu_k$,
	$\Hbar= \frac{1}{K}\sum_{k=1}^K H_k$.
	Then, up to the first order in $\epsilon$ the Sharpe ratio of the log-return of   $UP^2(\Mcal)$ is better (i.e., larger)
	than the Sharpe ratio of the log-return of  $UP^1(\Mcal)$. 
\label{prop:cont_time}
\end{proposition}

\bigskip
\noindent 
To ease notations we will write from now on sometimes $UP^\ell(T)$ instead of 
$UP^\ell(\Mcal)_T$.
In the proof of proposition \ref{prop:cont_time} we will need the following
technical lemmas, one that involves integration constants over $\Scal_K$ and another some properties of the variances of stochastic processes.

\begin{lemma}
Let $A_k=(A_k(t))_{t\ge0}$,  $B_k=(B_k(t))_{t\ge0}$, $k\le K$ be $L^2$ martingales starting from $0$. Denoting $\V$ the variance operator, the function~:
\begin{equation}
	x\mapsto \V\left[ \int_0^T \sum_{k=1}^K (1+x\cdot A_k(t))dB_k(t) \right]
	\label{eq:variance_increasing}
\end{equation}
is increasing with respect to $x$ for all $x\ge 0$. 
\label{lemma:variance_increasing}
\end{lemma}
\begin{proof}[{\bf Proof of lemma~\ref{lemma:variance_increasing}}]
The variance can be written as 
\begin{equation}
 \V\left[ \int_0^T \sum_{k=1}^K (1+x\cdot A_k(t))dB_k(t) \right] = \E \left[\int_0^T 
	\sum_{k,\ell=1}^K (1+ x A_k) \cdot (1+ x A_\ell) d\langle B_k,B_\ell\rangle_t \right].
\end{equation}
This is a second order polynomial in $x$; 
 the first order term is 
$ 2 \E \left[\int_0^T \sum_{k,\ell=1}^K A_k(t) d\langle B_k,B_\ell\rangle_t \right] = 
2 \int_0^T \sum_{k,\ell=1}^K \E \left[A_k(t)\right] d\langle B_k,B_\ell\rangle_t =0$;
the coefficient of $x^2$ is $\V\left[ \int_0^T \sum_{k=1}^K A_k(t)dB_k(t) \right]
$ thus positive.
The polynomial is therefore a constant plus $x^2$ times a positive coefficient hence the conclusion.
\end{proof}

\begin{lemma}
Suppose $K\ge 2$ and let $dw$ be the uniform measure  on the $K$ dimensional unit simplex denoted $\Scal_K$  defined in \eqref{eq:simplex}. 
Denote
\begin{equation}
	\alpha_K = \int_{\Scal_K} w_1 w_2 dw.
\end{equation}
Then for any $i,j \le K$~:
\begin{equation}
\int_{\Scal_K} w_i  dw = \frac{1}{K}, \ \ 
\alpha_K =  \frac{1}{K(K+1)}, \ \ 
\int_{\Scal_K} w_i w_j dw = \alpha_K (1+ \delta_{ij}).
\end{equation}
\label{lemma:simplex}
\end{lemma}
\begin{proof}[{\bf Proof of lemma~\ref{lemma:simplex}}]
	The uniform measure on the unit simplex is a particular case of Dirichlet distribution with all parameters equal to $1$; 
	the result is standard and follows from the formula of the moments of the Dirichlet distribution, see \cite{dirichlet_distrib2011}. For completeness we also give in Appendix~\ref{appendix:dirichlet} a self-contained proof.
\end{proof}
\begin{proof}[{\bf Proof of proposition~\ref{prop:cont_time}}]
Let us denote, for simplicity, $U=UP^1(\Mcal)$, $\Ucal=UP^2(\Mcal)$. From the previous relations and equation~\eqref{eq:up_as_average} we obtain~:
\begin{eqnarray}
	& \ &
U(T) = \int_{\Scal_K} exp\left({\sum_{k=1}^K  \left[\int_0^T w_k \mu_k^\epsilon(t) dt +
	w_k  H_k^\epsilon(T)\right] - \frac{1}{2}\sum_{k,\ell\le K} w_k w_\ell \int_0^T d\langle H_k^\epsilon , H_\ell^\epsilon \rangle_t  }\right)  dw
\nonumber \\ & \ &  
=e^{\int_0^T \muref(t)dt} \cdot
\int_{\Scal_K}
\left\{
 1+ \left(\sum_{k=1}^K  w_k \left[\int_0^T \epsilon \mu_k dt + \sqrt{\epsilon} H_k(T)\right] \right) \right.
\nonumber \\ & \ &  \left.
- \frac{\epsilon}{2} 
\sum_{k,\ell\le K} w_k w_\ell \int_0^T d\langle H_k , H_\ell \rangle_t + \frac{\epsilon}{2} \left(\sum_{k=1}^K  w_k H_k(T)\right)^2 \right\} dw + o(\epsilon)
\nonumber \\ & \ & 
=e^{\int_0^T \muref(t) dt} \cdot
\left\{1+ \int_0^T \epsilon \mubar(t) dt +  \sqrt{\epsilon} \Hbar(T)
 \right.
\nonumber \\ & \ & 
- \frac{\alpha_K\epsilon}{2} 
\sum_{k,\ell\le K} (1+\delta_{k\ell})\int_0^T d\langle H_k , H_\ell \rangle_t+ 
 \left.
\frac{\epsilon \alpha_K}{2} 
\sum_{k,\ell\le K}(1+\delta_{k\ell})H_k(T)H_\ell(T)
 \right\} + o(\epsilon). \label{eq:U1continuous}
\end{eqnarray}

The next steps of the proof are as follows: 

-- {\bf step A~:} use equation \eqref{eq:U1continuous}  in order to obtain the SDE for $\frac{dU}{U}$ 
up to order $1$ in $\epsilon$;

-- {\bf step B~:} adjoin $U$ to the market $\Mcal$ and do again a whole cycle of computations in order to obtain the log-dynamics $\frac{d \Ucal}{\Ucal}$; 

-- {\bf step C~:} compare $U$ and $\Ucal$ to obtain the insights stated in the proposition. 

\medskip \noindent
For step A we note that the term $e^{\int_0^T \muref(t) dt} $ is smooth so its quadratic covariation with the other terms is null and will just appear as an additive part in $dU/U$. On the other hand, $U = 1+\sqrt{\epsilon} \Hbar(T) + O(\epsilon)$ so
$1/U =  1-\sqrt{\epsilon} \Hbar(T) + O(\epsilon)$. We obtain
\begin{eqnarray}
	& \ & 
	\frac{dU}{U} = \muref(t) dt + \epsilon \mubar(t) dt + (1-\sqrt{\epsilon} \Hbar(t))\sqrt{\epsilon} d\Hbar(t)+
\nonumber \\ & \ &
+ \frac{\alpha_K\epsilon}{2} 
\sum_{k,\ell\le K} (1+\delta_{k\ell})[(-1)\cdot d\langle H_k , H_\ell \rangle_t
+d (H_k(t)H_\ell(t))]
+ o(\epsilon)
\nonumber \\ & \ &
=[\muref(t) + \epsilon \mubar(t)]dt+\sqrt{\epsilon} d\Hbar(t)
-\epsilon \Hbar(t) d\Hbar(t)+
\nonumber \\ & \ &
+ \frac{\alpha_K\epsilon}{2} 
\sum_{k,\ell\le K} (1+\delta_{k\ell})[H_k(t) d H_\ell(t) +  H_\ell(t) d H_k(t) ]
+ o(\epsilon)
\nonumber \\ & \ &
=[\muref(t) + \epsilon \mubar(t)]dt+\sqrt{\epsilon} d\Hbar(t)
\nonumber \\ & \ &
+ \epsilon \alpha_K  \left( \sum_{k=1}^K H_k(t) d H_k(t) - K \Hbar(t) d\Hbar(t) \right)
+ o(\epsilon)
\label{eq:dUoverUcontinuous0}
\end{eqnarray}
\noindent We obtain finally~:
\begin{eqnarray}
\!\!\!\!\!\!\
\frac{dU}{U} 
=(\muref+\epsilon \mubar)(t)dt+\sqrt{\epsilon} d\Hbar(t)
+ \epsilon \alpha_K  \left( \sum_{k=1}^K H_k d H_k(t) - K \Hbar d\Hbar(t) \right)
\!+\! o(\epsilon),
\label{eq:dUoverUcontinuous}
\end{eqnarray}
or, in equivalent form~:
\begin{eqnarray}
	\!\!\!\!\!\!\ 
	\boxed{
	\frac{dU}{U} 
	=(\muref+\epsilon \mubar)(t)dt+\sqrt{\epsilon} d\Hbar(t)
	+ \epsilon \alpha_K \sum_{k=1}^K (H_k-\Hbar) d H_k(t)
	\!+\! o(\epsilon).
	\label{eq:dUoverUcont_equiv}
}
\end{eqnarray}
\medskip \noindent
We continue with step B. The portfolio weights $w$ are now in $\Scal_{K+1}$ of which  the last coordinate $w_{K+1}$ indicates how much is allocated to $U$ (that has been added to the market $\Mcal$). As before~:
\begin{eqnarray}
	& \ &
	\Ucal(T) = e^{\int_0^T \muref(t)dt} \cdot 
	\int_{\Scal_{K+1}} exp\left\{  \sum_{k=1}^K  w_k \left[\int_0^T \epsilon \mu_k dt +
		\sqrt{\epsilon} H_k(T)\right] \right.
			\nonumber \\ & \ &  
		+ w_{K+1}\left[\int_0^T \epsilon \mubar (t) dt +\sqrt{\epsilon}\Hbar(T)\right] 
			+ \epsilon \alpha_K  w_{K+1} \int_0^T \sum_{k=1}^K (H_k-\Hbar) d H_k(t) 
			\nonumber \\ & \ &  		\left.
	-\frac{\epsilon}{2}\int_0^T d\left\langle w_{K+1}\Hbar+\sum_{k=1}^K w_k H_k ,w_{K+1}\Hbar+\sum_{\ell=1}^K w_\ell H_\ell \right\rangle_t dw+ o(\epsilon) \right\}
\nonumber \\ & \ &  			
= e^{\int_0^T \muref(t)dt} \cdot 
\int_{\Scal_{K+1}} exp\left\{  \sum_{k=1}^K  \tilde{w}_k \left[\int_0^T \epsilon \mu_k dt +
\sqrt{\epsilon} H_k(T)\right] \right.
\nonumber \\ & \ &  
+ \epsilon \alpha_K  w_{K+1} \int_0^T \sum_{k=1}^K (H_k-\Hbar) d H_k(t) 	
\nonumber \\ & \ &  		
\left.
-\frac{\epsilon}{2}\int_0^T d\left\langle \sum_{k=1}^K \tilde{w}_k H_k,\sum_{\ell=1}^K \tilde{w}_\ell H_\ell\right\rangle_t dw + o(\epsilon) \right\},
\end{eqnarray}
where we used the definition of $\mubar$ and $\Hbar$ and the notation
$\tilde{w}_k= w_k+\frac{w_{K+1}}{K}$.
Straightforward computations based on lemma~\ref{lemma:simplex} show that 
$\int_{\Scal_{K+1}} \tilde{w}_k dw = 1/(K+1)+ 1/K(K+1)= 1/K$
, $\int_{\Scal_{K+1}} \tilde{w}_k \tilde{w}_\ell dw =  \alpha_{K+1}(b_K + \delta_{k\ell})$  with $b_K = 1+2/K + 2/(K^2)$.
We continue and obtain, after some computations similar to the one before~:
\begin{eqnarray}
	& \ &
	\Ucal(T)  			
	= e^{\int_0^T \muref(t)dt} \cdot \left\{
1+	\left[\int_0^T \epsilon \mubar dt +
\sqrt{\epsilon} \Hbar(T)\right] 
	+ \frac{\epsilon \alpha_K}{K+1} \int_0^T \sum_{k=1}^K (H_k-\Hbar) d H_k(t)
\right.
	\nonumber \\ & \ &
\left. 
	+\epsilon \alpha_{K+1} \int_0^T \left( \sum_{k,\ell=1}^K  (b_K+\delta_{k\ell}) (H_k-\Hbar) d (H_\ell -\Hbar) \right) 
o(\epsilon) \right\}.
\end{eqnarray}
We continue by computing $d\Ucal / \Ucal$ to the first order in $\epsilon$. Again, the computations are not difficult but rather lengthy and we only give the result~:
\begin{eqnarray}
	\!\!\!\!\!\!\
\boxed{
	\frac{d\Ucal}{\Ucal} 
	=(\muref+\epsilon \mubar)(t)dt+\sqrt{\epsilon} d\Hbar(t)
	+ \epsilon \alpha_{K}(1-K \alpha_{K+1})\cdot \left( \sum_{k=1}^K (H_k-\Hbar) d H_k(t)\right)
	\!+\! o(\epsilon).}
	\label{eq:dU2overU2cont_equiv}
\end{eqnarray}
This equation looks like \eqref{eq:dUoverUcont_equiv} with the only exception that the last term has a smaller coefficient~: $\alpha_{K}(1-K\alpha_{K+1})$  instead of $\alpha_{K}$. 
 This remark is exploited in the step C of our proof to which we turn now.

\medskip \noindent 
 We will work with the log-returns associated with $U$ and $\Ucal$. For a strictly positive It\^o process $Y$ with $dY/Y = a dt + b dL_t$ 
with $L$ a 
 martingale starting from $0$ (and under suitable regularity assumptions),
 $\ln(Y(T)/Y(0)) = \int_0^T a+\frac{b^2}{2}d\langle L,L\rangle_t+ b dL_t$ which means that the mean is 
 $\E  \left[\int_0^T a+\frac{d\langle L,L\rangle_t}{2}dt\right]$ and the standard deviation 
 $\sqrt{\E \left[ \int_0^T b^2 d \langle L,L\rangle_t\right]}$. 
 So here, to the first order in $\epsilon$ the means of 
  $\ln(U(T)/U(0))$ and  $\ln(\Ucal(T)/\Ucal(0))$ are the same
\begin{eqnarray}
& \  & 
\E \left[\ln(U(T)/U(0))\right]
 =\E \left[ \int_0^T (\muref(t) + \epsilon\mubar)  dt+ \epsilon \int_0^T d\langle \Hbar,\Hbar\rangle_t \right] + o(\epsilon) \nonumber
\\ & \ & 
= \E \left[\ln(\Ucal(T)/\Ucal(0))\right]+ o(\epsilon).
\end{eqnarray}
Therefore, the numerators in the Sharpe ratios for $U$ and $\Ucal$ are of order $O(1)$ and only differ at order $o(\epsilon)$, so are equal to leading order. 
On the other hand note that the stochastic part of $UP^\ell(\Mcal)$ ($\ell=1,2$) can be written 
as $\frac{\sqrt{\epsilon}}{K}\sum_k (1 + c_\ell (H_k-\Hbar))d H_k$ for some constants $c_1$ and $c_2$ with $c_2 < c_1$ (constant $c_\ell$ corresponds to  $UP^\ell(\Mcal)$). 
Then from the lemma~\ref{lemma:variance_increasing}
it follows that the variance hence the standard deviation of  $\ln(\Ucal(T)/\Ucal(0))$ is smaller than that $\ln(U(T)/U(0))$~:
\begin{eqnarray}
	std \left[\ln(\Ucal(T)/\Ucal(0))\right] \le	std \left[\ln(U(T)/U(0))\right] + o(\epsilon).
\end{eqnarray}
In the Shape ratios of $U$ and $\Ucal$, $std \left[\ln(\Ucal(T)/\Ucal(0))\right]$ and	$std \left[\ln(U(T)/U(0))\right]$ appear as   denominators; both are of order $\sqrt{\epsilon}$ and $std \left[\ln(U(T)/U(0))\right]$ is significantly larger  than 
 $std \left[\ln(\Ucal(T)/\Ucal(0))\right]$ because their difference is a positive term of order $\sqrt{\epsilon}$ plus some unknown error of order $o(\epsilon)$. So the denominator for 
$\Ucal$ is significantly smaller than that for $U$ and 
thus the Sharpe ratio of $\Ucal$ is larger than that of $U$ which ends the proof.
\end{proof}
}

\section{Numerical results and discussion}\label{sec:numerical}

\subsection{Computation algorithm}

Although some algorithms exists to compute the Cover UP using combinatorial identities, see \cite{cover1996universal} for the case of the Dirichlet(1/2,...,1/2) distribution, they are known to 
feature, in full rigor, exponential complexity
with respect to
the dimension $K$. 
 On the other hand, Monte Carlo approaches have been shown to perform well, see 
\cite{blum1997universal,ishijima_numerical_2001} and we will take this choice here. Note also the special 
sampling like in \cite{kalai2002efficient} that obtain polynomial complexity 
 when sampling from a non-uniform distribution.

 We resort then to a straightforward loop over $t\in \Tcal$ by evaluating the simplex averages in \eqref{eq:defup}. Therefore, to compute Cover UP~: 
\begin{itemize}
	\item when there are exactly two assets the expectation with respect to the uniform law over the simplex $\Scal_2$ can be computed as an integral over $x\in [0,1]$ taking $w=(x,1-x) \in \Scal_2$. In this case
	we use a $16$ points Gauss-Legendre quadrature over the interval $[0,1]$.
	Denote $q_j, \xi_j$, $j=1,...,16$ the weights and points of the Gauss-Legendre quadrature; recall that $\xi_j \in [-1,1]$ and the quadrature is designed for functions defined over this interval; to compute the averages over $\Scal_2$ we use, for any function $\Hcal:\Scal_2 \to \R$ the formula~: 
	\begin{equation}
		\E_{(b_1,b_2)\in \Scal_2} [\Hcal(b_1,b_2)]= \int_0^1 \Hcal(x,1-x)dx \simeq \frac{1}{2} \sum_{j=1}^{16} q_j \Hcal\left( \frac{1+\xi_j}{2}, \frac{1-\xi_j}{2} \right).
	\end{equation}
	\item when there are more than $3$ assets we draw  $M=10'000$ random points\footnote{We chose  $M=10'000$ because empirically the results obtained are stable for this value while still being tractable in a resonable amount of time}  
	$\xi_1, ..., \xi_M \in \Scal_K$ 
	from the unit simplex and replace the exact expectation with a sample average~: 
	\begin{equation}
		\E_{b\in \Scal_K} [\Hcal(b)] \simeq \frac{1}{M} \sum_{j=1}^{M} \Hcal\left( \xi_j \right). \label{eq:MonteCarloapprox}
	\end{equation}
	The points $\xi_j$ are drawn using the property that if $X_1,...,X_K$ are independent exponentially distributed variables then $\left(\frac{X_1}{\sum_j X_j}, ..., \frac{X_K}{\sum_j X_j}\right)$ follows the uniform law on $\Scal_K$. 

Together with relation \eqref{eq:up_as_average} this allows to compute $UP^1_t$ for all $t\le T$. Once this first step done, we add $UP^1$ to the market $\Mcal$  as 
$f^{K+1}_t= UP^1_t/ UP^1_{t-1}$
and proceed recursively with the computation of all other $UP^\ell$ for $\ell>1$. If we use $n_S=10'000$ samples to compute the integral over the simplex, the complexity of an algorithm to compute $UP^\ell_t$ for $\ell \le L$, $t\le T$ will be at most 
$O(n_S \cdot T \cdot \left(K+\frac{L}{2}\right) \cdot L )$.
	
\end{itemize}

\subsection{Toy example from the literature}
\label{sec:cst_plus_oscillating}

\begin{figure}[htpb!]
	\centering
	\includegraphics[width=.9\linewidth,height=0.3\textheight]{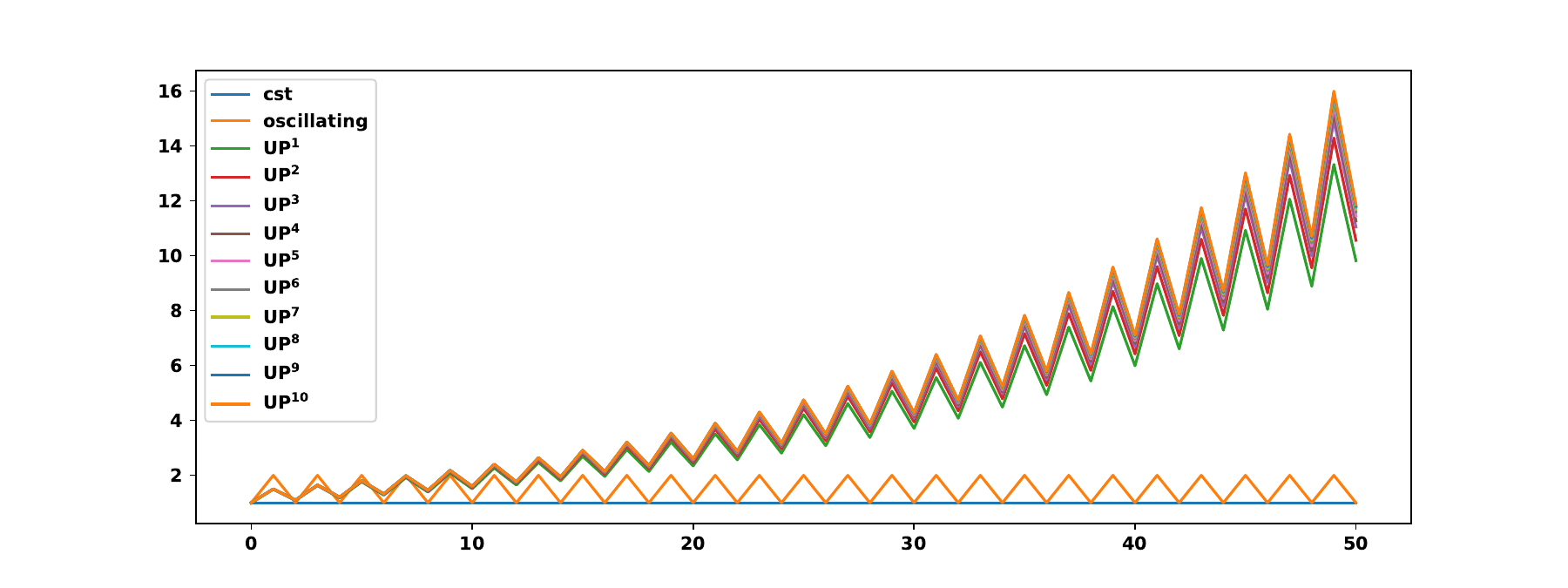}
	
	\includegraphics[width=.9\linewidth,height=0.3\textheight]{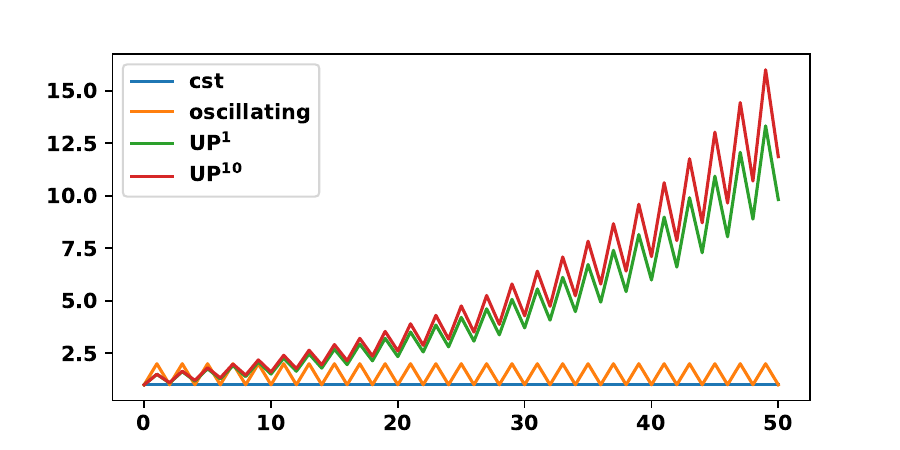}
	
	\includegraphics[width=.9\linewidth,height=0.3\textheight]{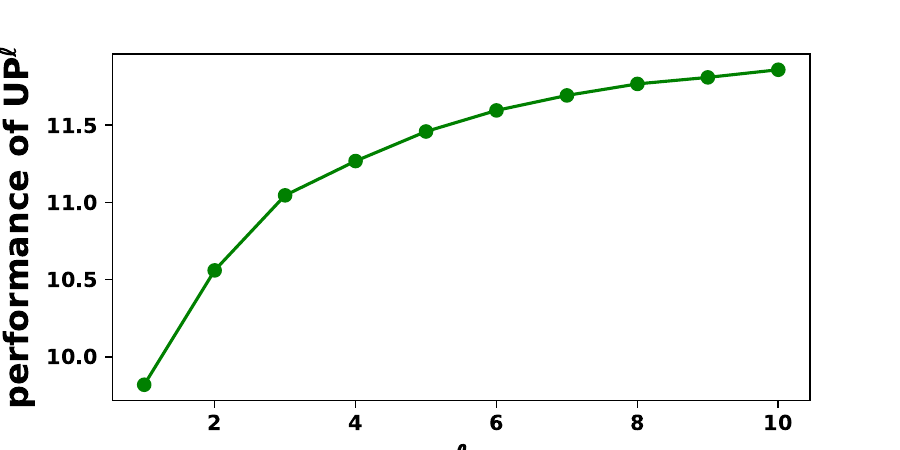}
	\caption{The result of the high order universal portfolios for the 
		toy example in section~\ref{sec:cst_plus_oscillating}.
{\bf Top~:} the performance of the two assets and of the first $10$ high order universal portfolios ($UP^1$ being the classical, Cover, UP).  {\bf Middle~:} for convenience only Cover ($UP^1$) and $UP^{10}$ (the last) portfolios are plotted together with the individual assets. {\bf Bottom~:} the performance of $UP^\ell$ depending on $\ell$ is seen to  increase with $\ell$.
	}\label{fig:cover2assets}
\end{figure}
We start with a toy example taken from \cite[Section V]{cover1996universal}, see also \cite{kalai2002efficient}  : one asset is  constant, i.e. all price relatives equal $1$, and the other has oscillating behavior: price relatives are $2$, then $1/2$ then $2$ again, i.e. returns to the initial value after any even number of time steps. We set $T=50$.
The results presented in Figure~\ref{fig:cover2assets} confirm the basic properties of the Cover UP and the remark~\ref{rem:optimality}: the UP improves over the individual performances and HOUP improves when compared to the UP.

\subsection{'Old NYSE' dataset}
\label{sec:numerical_nyse_o}

We implemented the high order universal portfolios on data from \cite{cover91} which contains $22$ years of daily quotations from $1965$ to $1987$ ($T=5650$) for $K=36$ stock symbols quoted on the New York Stock Exchange. This dataset was considered by all subsequent researchers and allows to compare results with the original proposal. To obtain the dataset the reader can consult \cite{Helmbold98}, \cite{kalai2002efficient} or the github package
\cite{marigold_github_up} (the dataset corresponds to the ``nyse\_o.csv'' file).

In all cases we plot the comparison between the asset dynamics and the values of the $UP^\ell$ for $\ell$ up to $10$; as Cover, we use pairs of two assets, see 
 Table \ref{table:nyse_o_couples}  for a qualitative description of each pair  reproduced from
\cite[p. 122]{dochow_proposed_2016}.

\begin{table}[h]
\caption{Descriptions of the markets considered in section~\ref{sec:numerical_nyse_o}.}\label{table:nyse_o_couples}%
\begin{tabular}{@{}c|l|l|l|l@{}}
\toprule
Market & Asset & Correlation & individual& Description  \\
number&  names &            & performances& cf. \cite{dochow_proposed_2016} \\
\midrule
$1$ & Commercial Metals   & 0.064& 52.02   & Volatile and stagnant\\
& Kin Ark             &      & 4.13  & uncorrelated\\
\botrule
$2$ & Irocquois           & 0.041& 8.92   & Volatile \\
    & Kin Ark             &      & 4.13  & uncorrelated\\
\botrule
$3$ & Coca Cola       & 0.388& 13.36   & Non-volatile \\
    & IBM             &      & 12.21  & highly correlated\\
\botrule
$4$ & Commercial Metals   & 0.067& 52.02   & Volatile\\
&  Meicco             &      & 22.92  & uncorrelated\\
\botrule
\end{tabular}
\end{table}

\begin{figure}[htpb!]
	\centering
	\includegraphics[width=.9\linewidth,height=0.3\textheight]{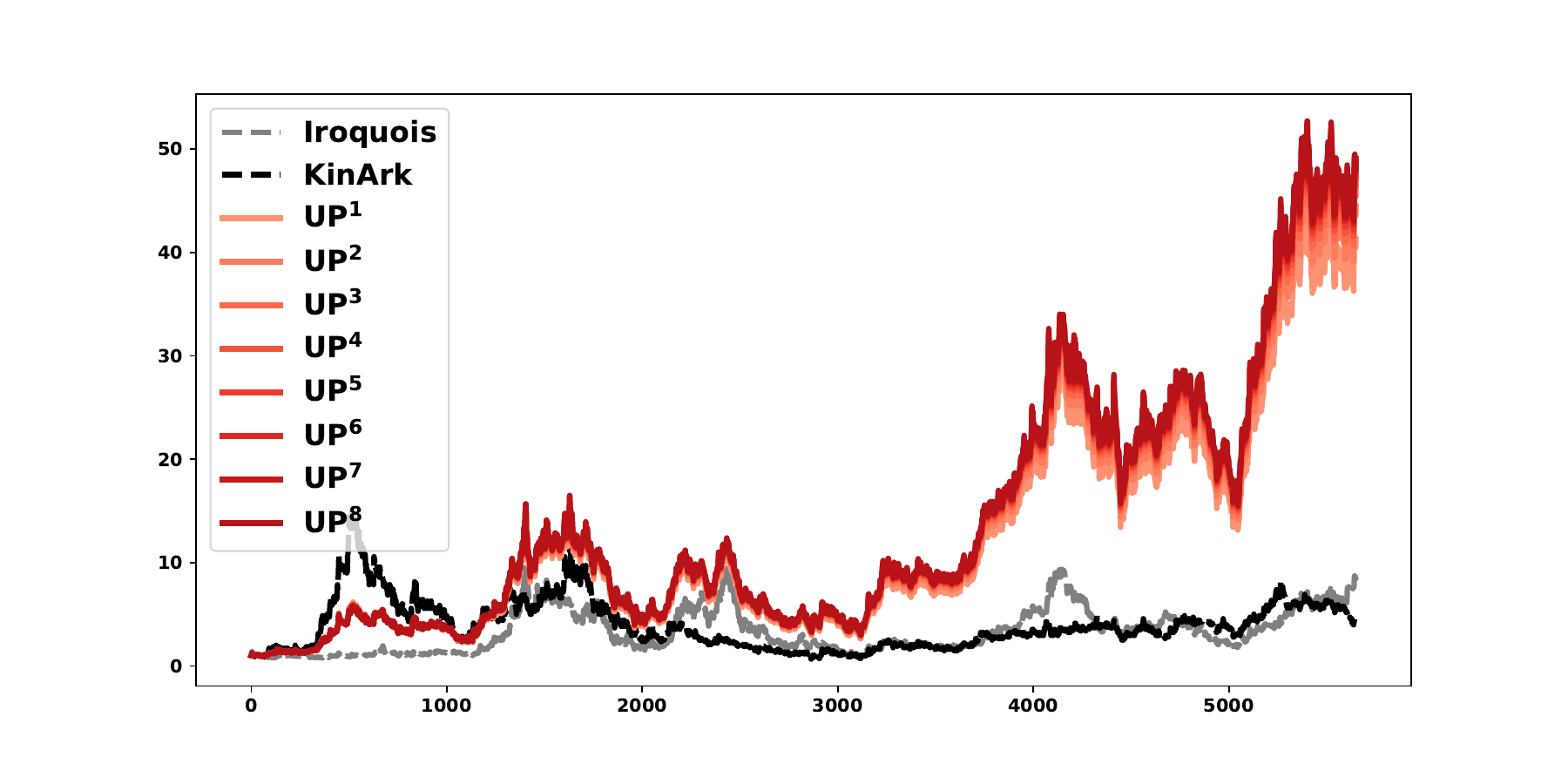}

	\includegraphics[width=.9\linewidth,height=0.3\textheight]{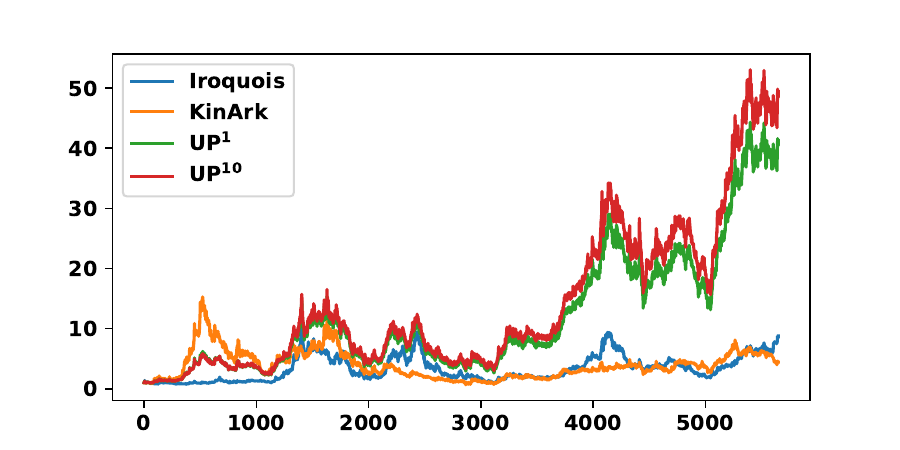}

	\includegraphics[width=.9\linewidth,height=0.3\textheight]{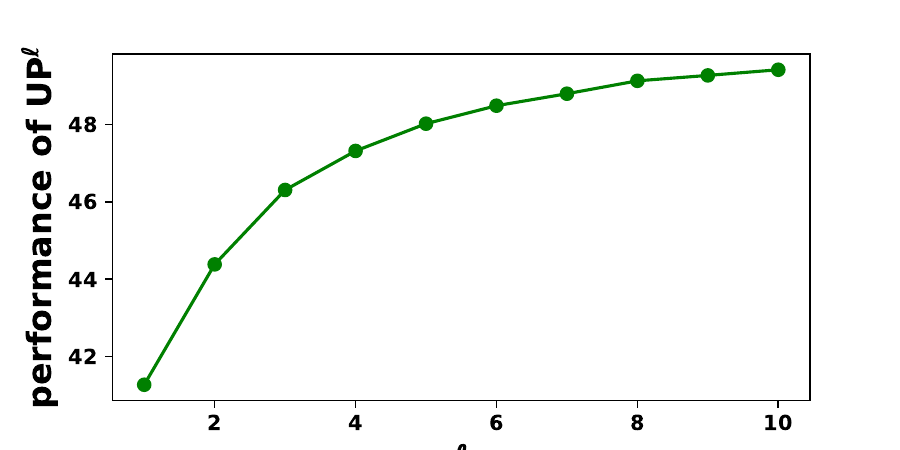}
	\caption{The result of the high order universal portfolios for the couple ``Iroquois''-``Kin Ark''. {\bf Top~:} the performance of the two assets and of the first $10$ high order universal portfolios ($UP^1$ being the classical, Cover, UP). It is seen that the performance is, as expected, above that of individual assets. {\bf Middle~:} for convenience only Cover ($UP^1$) and $UP^{10}$ (the last) portfolios are plotted together with the individual assets. {\bf Bottom~:} the performance of $UP^\ell$ depending on $\ell$ which is seen to  be increasing with $\ell$.
	}\label{fig:ir_ka}
\end{figure}

\begin{figure}[htpb!]
\includegraphics[width=.9\linewidth,height=0.3\textheight]{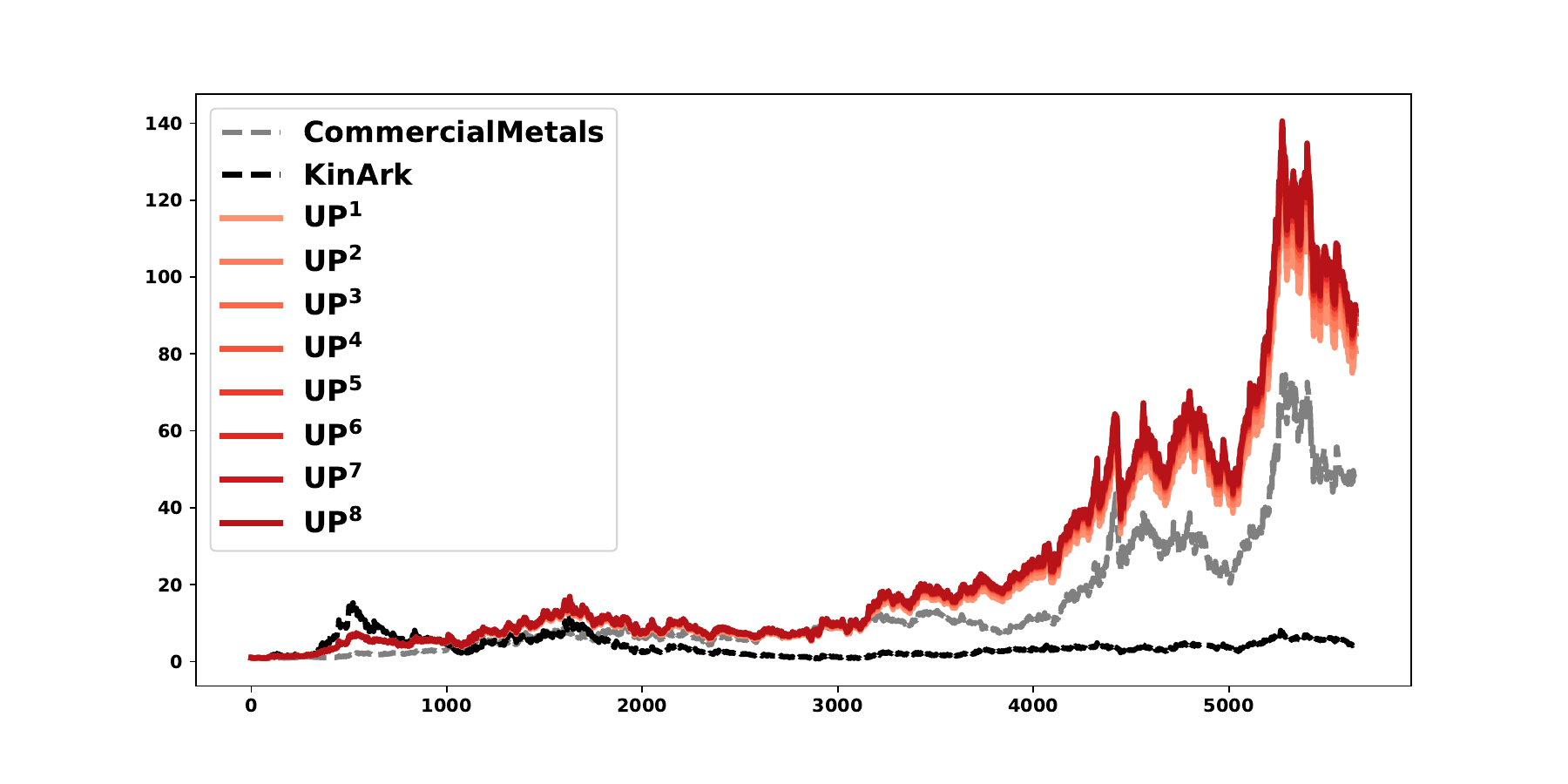}

\includegraphics[width=.9\linewidth,height=0.3\textheight]{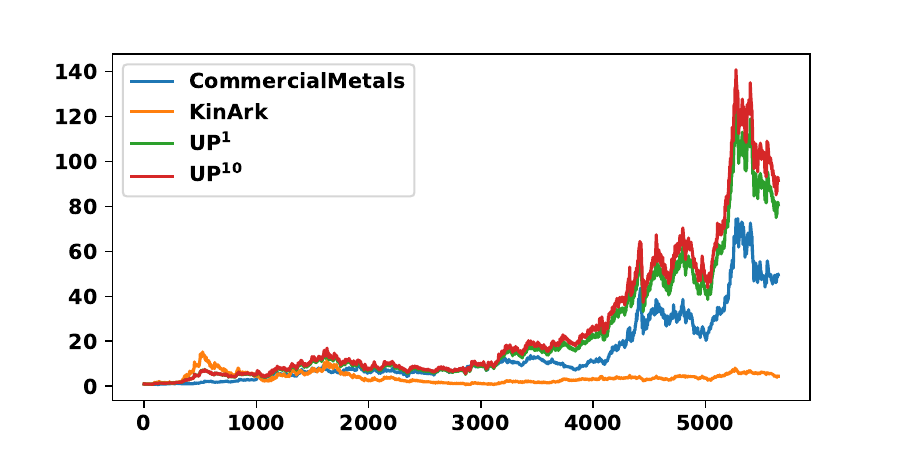}

\includegraphics[width=.9\linewidth,height=0.3\textheight]{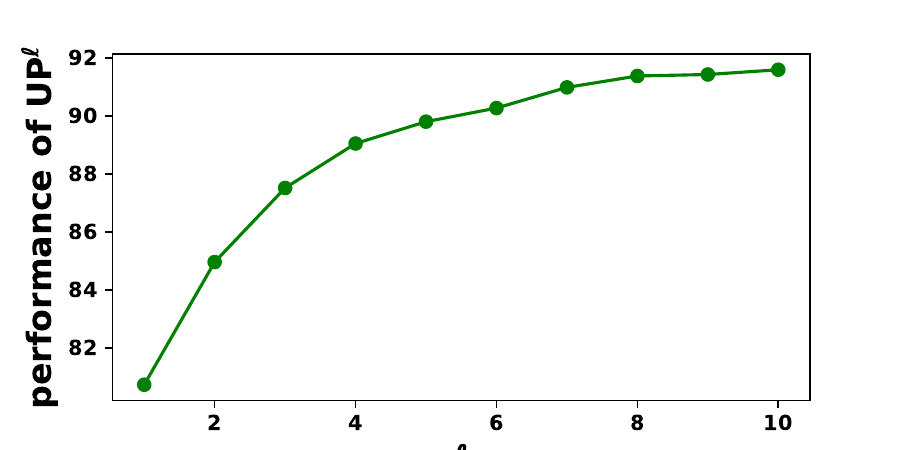}
\caption{Same as in figure \ref{fig:ir_ka} for the couple ``Commercial Metals''-``Kin Ark''.}\label{fig:com_ka}
\end{figure}

\begin{figure}[htpb!]
	
	\includegraphics[width=.9\linewidth,height=0.3\textheight]{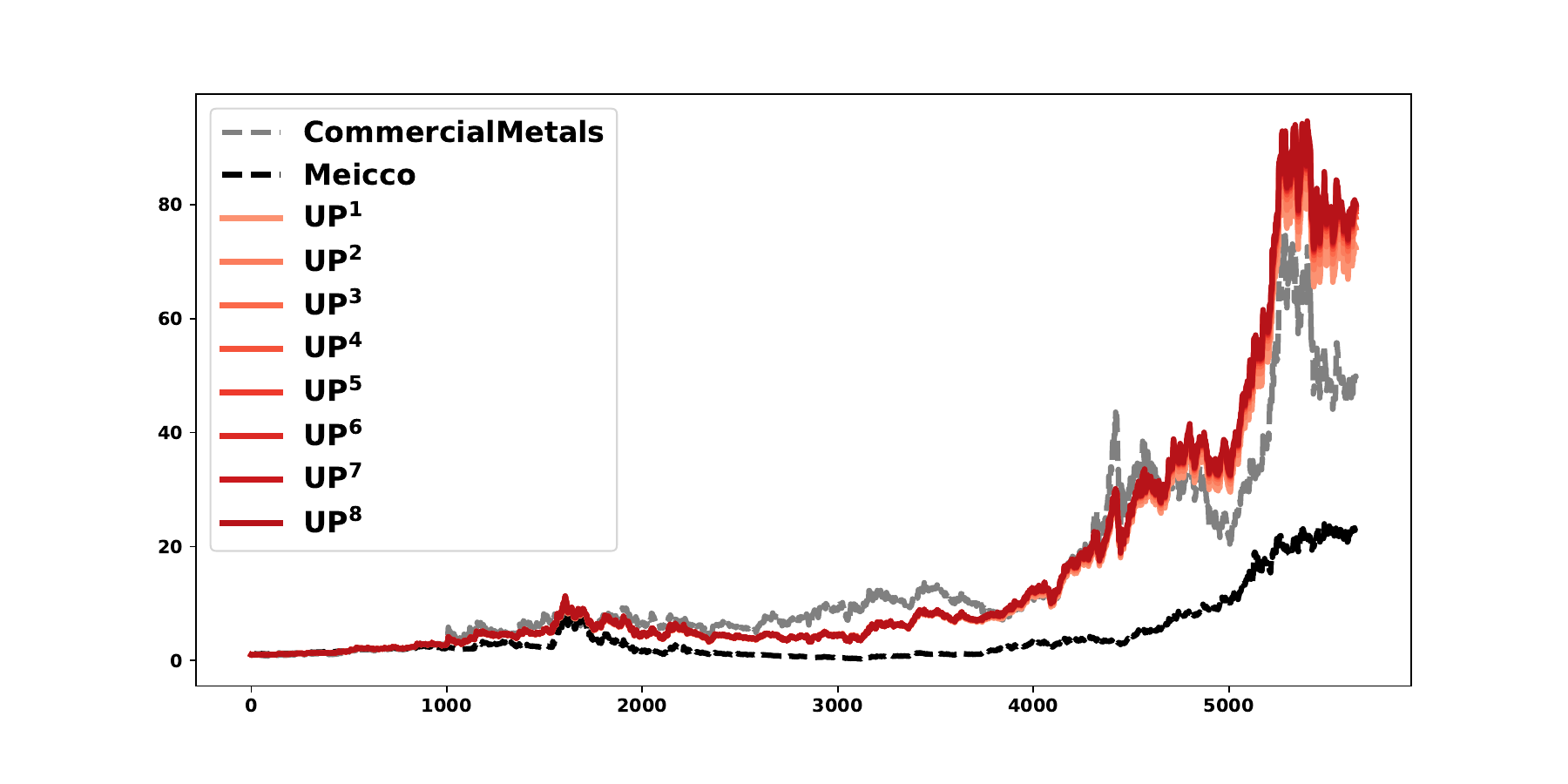}
	
	\includegraphics[width=.9\linewidth,height=0.3\textheight]{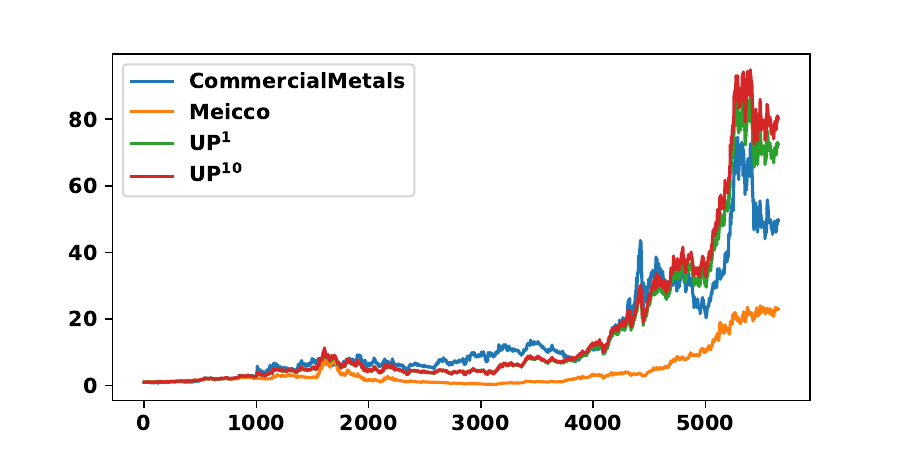}
	
	\includegraphics[width=.9\linewidth,height=0.3\textheight]{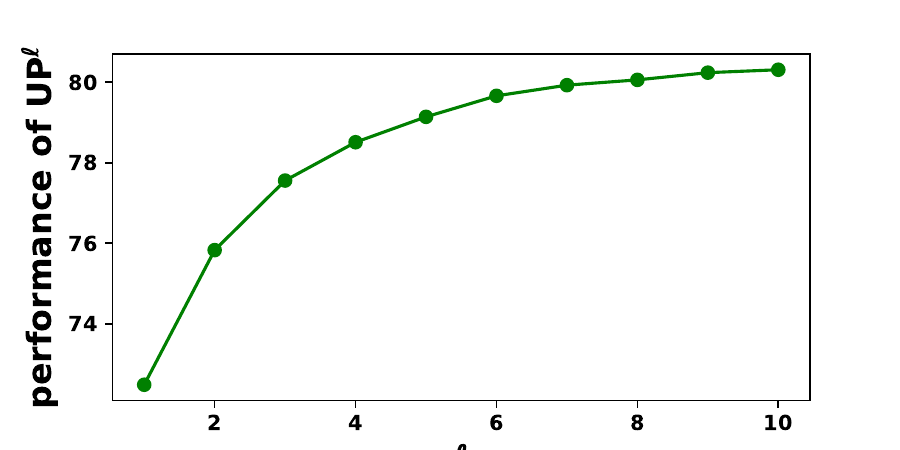}
	\caption{Same as in figure \ref{fig:ir_ka} for the market $4$ in table~\ref{table:nyse_o_couples} :  ``Commercial Metals''-``Meicco''.}\label{fig:cm_mei}
\end{figure}

\begin{figure}[htpb!]

\includegraphics[width=.9\linewidth,height=0.3\textheight]{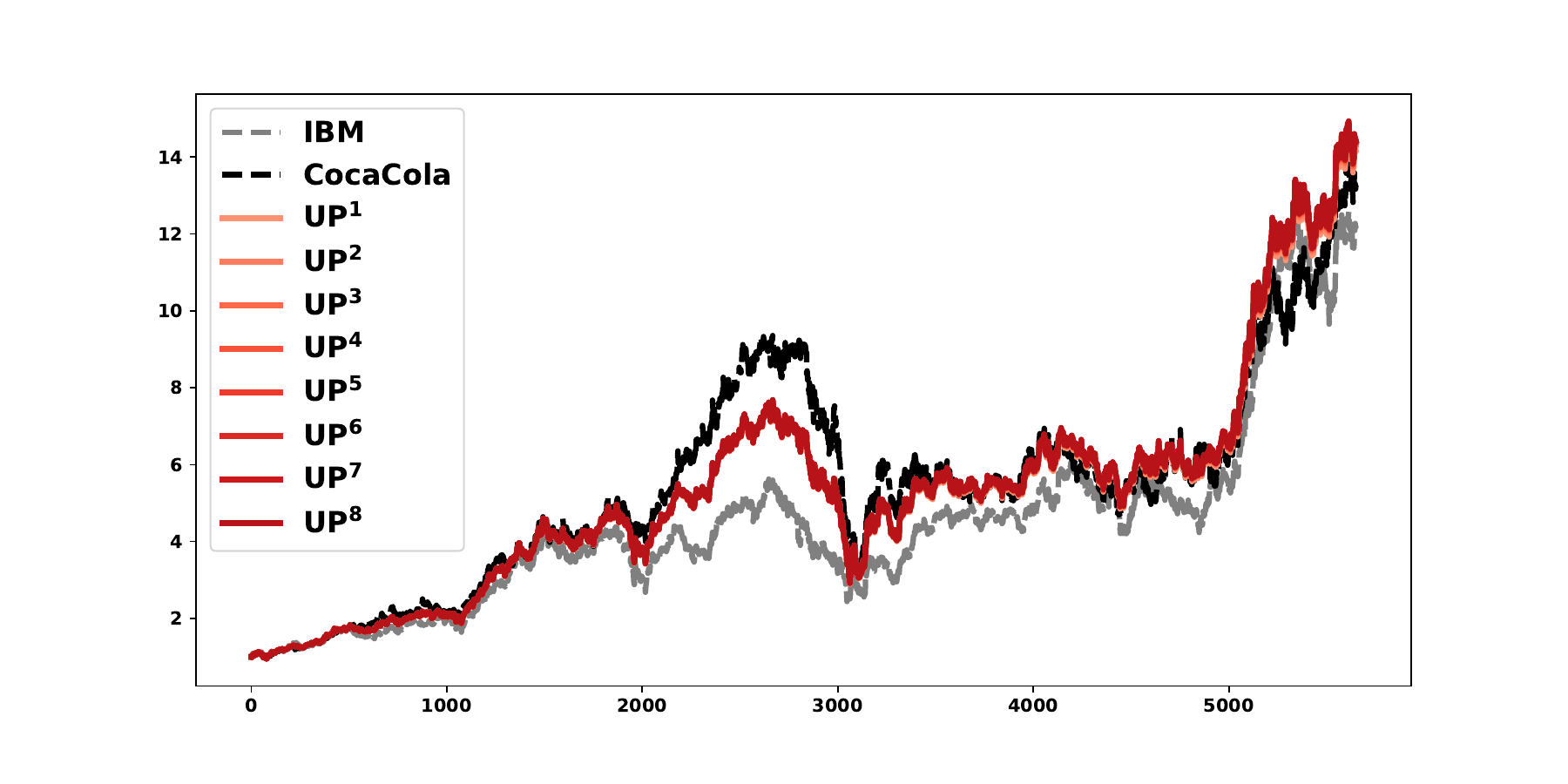}

\includegraphics[width=.9\linewidth,height=0.3\textheight]{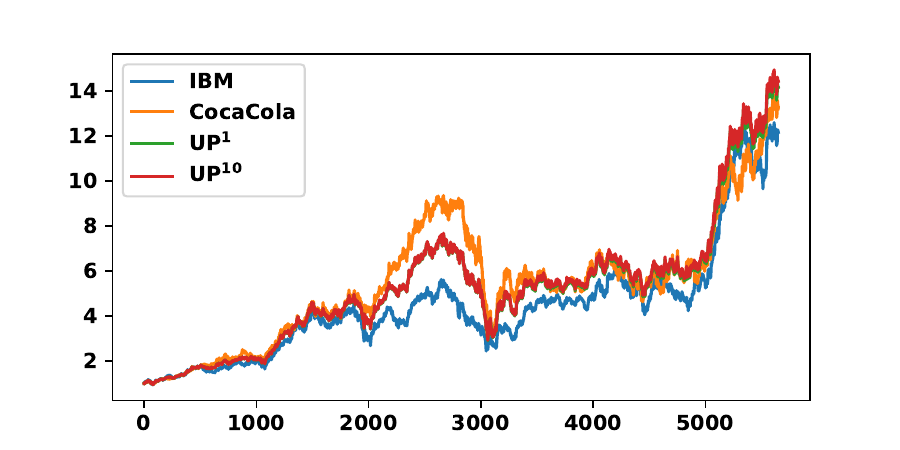}

\includegraphics[height=0.3\textheight]{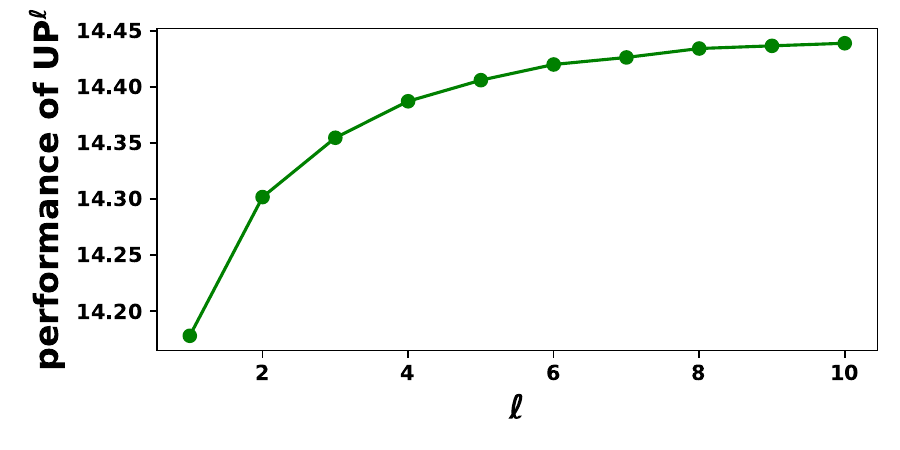}
\caption{Same as in figure \ref{fig:ir_ka} for the couple ``IBM''-``Coca Cola''.}\label{fig:ibm_cc}
\end{figure}

We also plot the dependence of the final value of the $UP^\ell$ with respect to $\ell$ to check the remark~\ref{rem:optimality}.
The results are presented in figures \ref{fig:ir_ka}-\ref{fig:ibm_cc}. 
First of all it is noted that in general the performance of the $UP^\ell$ portfolios
is evolving gradually with $\ell$ with no severe discontinuities. On the other hand the performance appears to be always increasing with $\ell$, in agreement with remark~\ref{rem:optimality}.

For instance, a situation that is known to provide good results (cf. \cite{cover91}) is the couple ``Iroquois'' -- ``Kin Ark''  (figure \ref{fig:ir_ka}). In this case the best
individual performance is around $8$ times the initial wealth, while the Cover UP provides around $40$ times the initial wealth over the $22$ year period. But, letting $\ell$ increase, $UP^\ell$ manages to extract even more additional performance with $UP^{10}$ reaching a further $20\%$ over Cover's UP ($49$ vs $41$) and in any case well above individual performances. Given the form of the curve one can expect that the limit $UP^\ell$ when $\ell$ increases will 
stabilize around $50$ and this is indeed what numerical tests (not plotted here) show.

Another benchmark test is the couple ``Commercial Metals'' -- ``Kin Ark'' (figure \ref{fig:com_ka}). Same favorable behavior is obtained, with performance of Cover UP being around $80$ and $UP^{10}$ at about $90$, compared with best individual performance at cca. $50$ times initial wealth.

We also tested the couple ``Commercial Metals'' -- ``Meicco'' (figure \ref{fig:cm_mei}). Again, the performance of the Cover UP of around being around $72$ is improved by $UP^{10}$ to cca. $80$; here the best individual performance is situated at $~50$ times initial wealth.

Finally, we consider a situation where it is known that the UP will not work better than the individual assets, the ``IBM'' -- ``Coca Cola'' couple. Here all $UP^\ell$ appear very close to $UP^1$ the good news being that there is no degradation in the performance when $\ell$ increases which still remains of the same order of the best individual asset (at about $13$ times initial wealth).

{
\subsection{Further numerical tests}

\subsubsection{Larger markets} \label{sec:perf5market}

We consider now a situation with more than two assets. We considered five assets of the ``Old NYSE'' dataset and  took at random $1000$ samples of $5$-tuples from the initial market and then computed the high order $UP$s on this set. Each of the $1000$ experiments uses a {\bf different} $5$-tuple of assets sampled randomly from the $\frac{1000!}{5! \cdot 995!}= 8'250'291'250'200$ possible choices. The purpose of this test is to assess that the $UP^\ell$ portfolios are indeed useful i.e. to compare the performance of $UP^\ell$ for $\ell >1$ with the standard Cover universal portfolio $UP^1$.
The results are in figure \ref{fig:confidence_intervals_5tuples_up110}. The $1000$ markets are in yellow and the quantiles of the distribution in different colors. 
We plot the quotient  $UP^\ell(T)/UP^1(T)$. 
It is seen that in over $90\%$ cases $UP^\ell$ improves over $UP^1$ and in the upper decile the relative improvement is over $5\%$. A non-parametric Wilcoxon test confirmed that there is a very low likelihood of the data not being significant 
(p-value for $UP^1$ not being lower than $UP^{10}$ is evaluated at $4.7\cdot 10^{-159}$.)
We conclude that $UP^\ell$ does improve over $UP^1$.
Even if this difference may seem small it could constitute an arbitrage opportunity as indicated in figure 
\ref{fig:confidence_intervals_5tuples_up110_arbitrage} where we plot (in $\%$) the
additional performance of $UP^{10}$ vs. $UP^{1}$ i.e., the quotient $\frac{UP^{10}(T)-UP^1(T)}{UP^1(T)}$. The $10$-th order universal portfolio  $UP^{10}$ appears to consistently outperform $UP^1$~; this means that a portfolio that is long $UP^{10}$ and short $UP^1$ is often profitable~\footnote{Disclaimer~: of course, this is valid in this particular dataset and does not constitute a guarantee that it will be true in other situations; further empirical tests should be performed before any real-life investment decisions.}.

\begin{figure}[htpb!]
	\centering
	\includegraphics[width=0.95\linewidth]{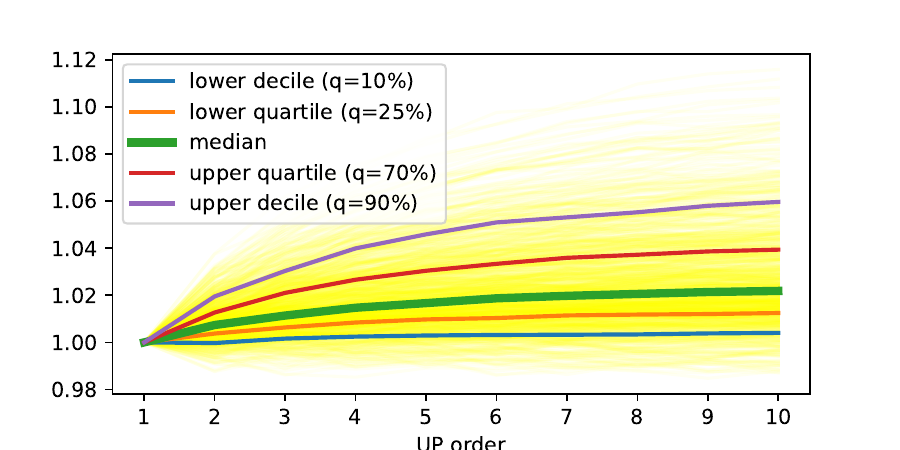}
	\caption{Quotient $UP^\ell(T)/UP^1(T)$ for $1000$ markets in 
		section \ref{sec:perf5market}
		consisting of $5$ assets taken randomly from the 'Old NYSE' dataset. The $1000$ markets are in yellow, the quantiles in other colors as indicated in the legend.  See also
		figure 	\ref{fig:confidence_intervals_5tuples_up110_arbitrage} for an 'arbitrage view'. 
	}\label{fig:confidence_intervals_5tuples_up110}
\end{figure}

\begin{figure}[htpb!]
	\centering
	\includegraphics[width=0.4\linewidth]{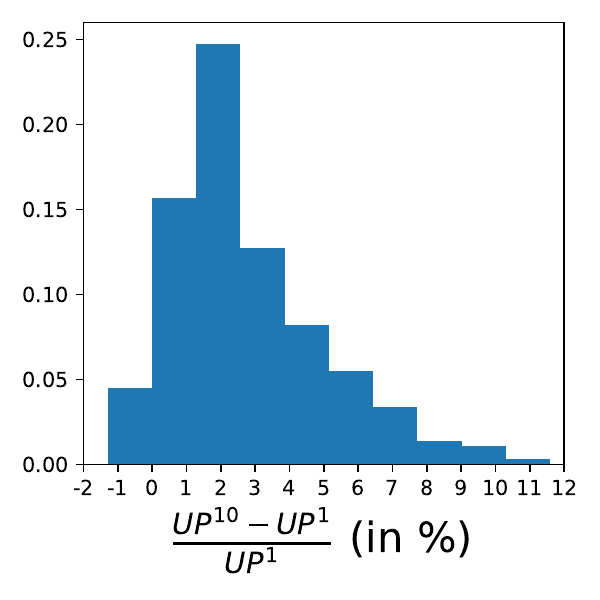}
	\caption{Arbitrage test for $UP^{10}$ versus $UP^1$, cf. also
		figure \ref{fig:confidence_intervals_5tuples_up110_arbitrage}.
We consider $1000$ markets of $5$ assets each cf. section \ref{sec:perf5market}.}
	\label{fig:confidence_intervals_5tuples_up110_arbitrage}
\end{figure}

Although not the primary goal of our contribution, we also compared the (high order)   universal portfolios introduced here with the ``Buy\&Hold'' strategy. The results in figure \ref{fig:confidence_intervals_5tuples_buy_hold} indicate that in more than $90\%$ of the situations the (high order) universal portfolios outperform the ``Buy\&Hold'' strategy~; in the $25\%$ best cases (upper quartile) the final value of the universal portfolios are about the double of the ``Buy\&Hold'' strategy and in $10\%$ best 
 more than triple. The  best sample gains $+682\%$ while the worse sample loses 
$18\%$ with respect to the ``Buy\&Hold'' reference.

\begin{figure}[htpb!]
\centering
\includegraphics[width=0.85\linewidth]{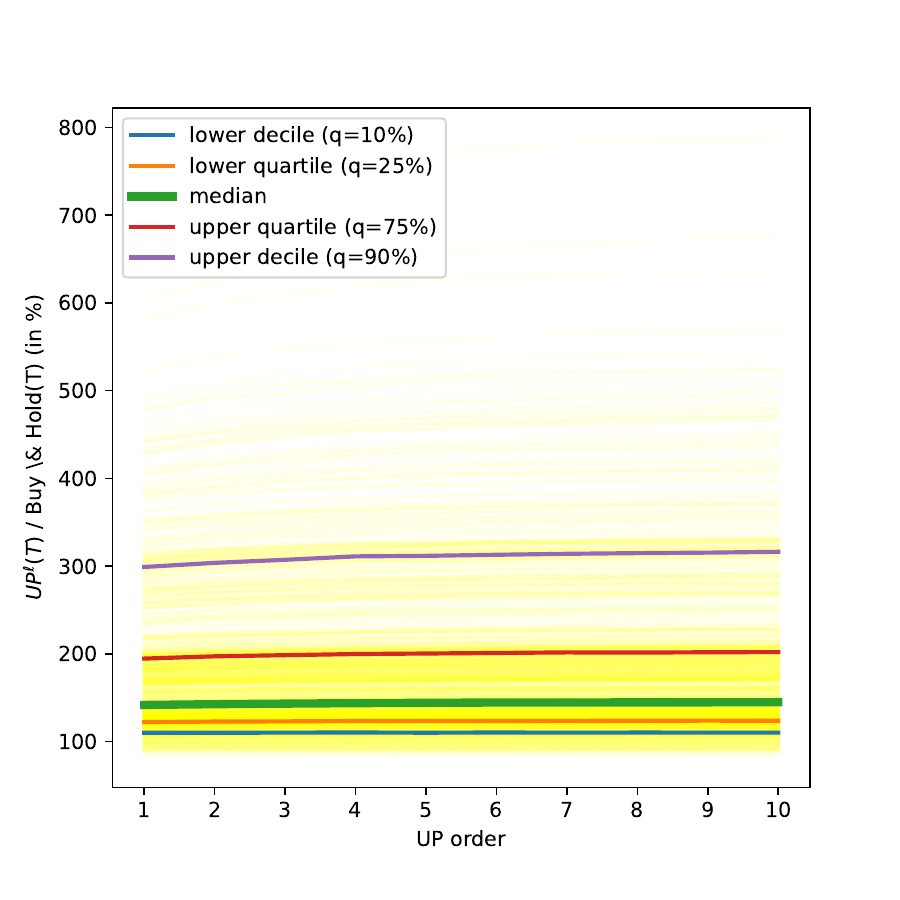}
\caption{Performance of $UP^\ell$ relative to a ``Buy\&Hold'' strategy for the test case in section~\ref{sec:perf5market}. The $1000$ tests are in yellow, quantiles in other colors as indicated in the legend. 
}\label{fig:confidence_intervals_5tuples_buy_hold}
\end{figure}

\subsubsection{Confidence intervals}	\label{sec:confidence_intervals}

	To further assess the quality of the conclusions above we investigate the approximation used to compute the $UP^\ell$ values. We fix arbitrarily a market consisting of the first five assets of the ``Old NYSE'' dataset i.e. columns $1$ to $5$ of the dataset (other choices give very similar results), see section ~\ref{sec:numerical_nyse_o}. 
	The main approximation \eqref{eq:MonteCarloapprox} consists in replacing the integration over the simplex $\Scal_K$ by an empirical average over $10'000$ points drawn uniformly from the simplex. So in fact we are computing a Monte Carlo average.
We take $10'000$ samples from the unit simplex $\Scal_K$  and compute the performance of the $UP^\ell$ portfolios for $\ell=1,..., 10$. We do this $1'000$ times 
	and plot the statistics for the performance $UP^\ell(T)$ at the final time.
	 This is a standard procedure for testing a Monte Carlo approximation. Ideally all $1000$ samples should give the same result but this will not happen because of the Monte Carlo procedure is only an approximation of the integral.
	
	In this case the standard ``Buy\&Hold'' strategy gives a performance of $18.66$ (independent of any sampling).
	The results are presented in figure \ref{fig:confidence_intervals} which shows that, even if there is considerable noise in the computation, the values are still stable enough to be confident about the conclusions obtained.
	
	In addition we used a non-parametric Wilcoxon test to inquire whether $UP^1 \le UP^{10}$ i.e., we tested it they could be equally distributed. Up to a p-value of $1.7\cdot 10^{-165}$ this does not seem to be validated by the data so one concludes, as the right plot in figure 
	\ref{fig:confidence_intervals} indicates, that $UP^1(T) \le UP^{10}(T)$.
	Same test indicate that one should reject the hypothesis that both are comparable to the ``Buy\&Hold'' performance and indicate instead that both are significantly larger than the 
	``Buy\&Hold'' performance (this seems obvious from the left plot in  figure 
	\ref{fig:confidence_intervals}).
	
	
	\begin{figure}[htpb!]
		\centering
		\includegraphics[width=0.65\linewidth]{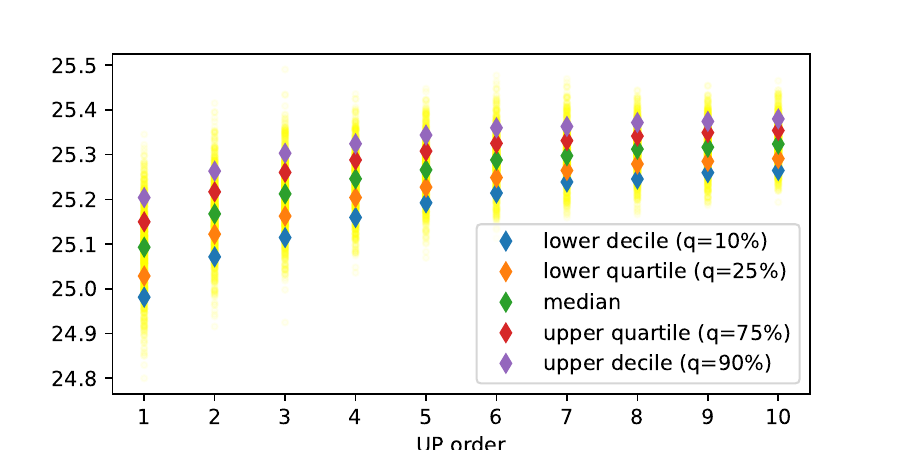}
		\includegraphics[width=0.33\linewidth]{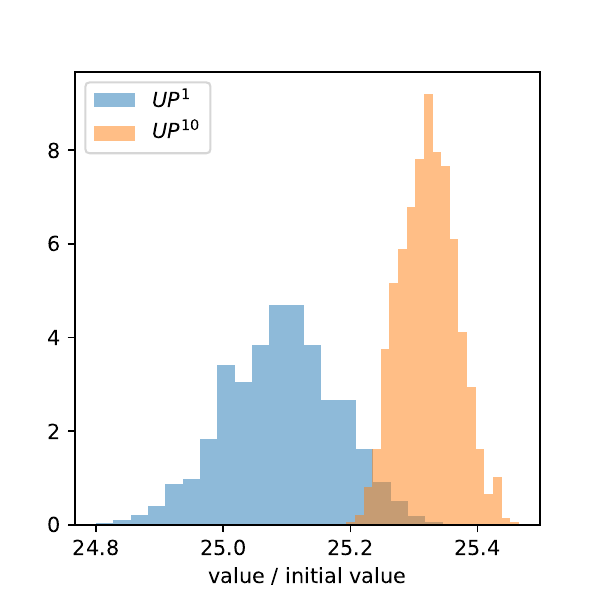}
		\caption{Statistics of the computation of $UP^\ell$, cf. section \ref{sec:confidence_intervals}. The same market is used here but we take $1'000$ distinct simulations, each including $10'000$ samples of the unit simplex. {\bf Left:} the results for all $UP^\ell$ portfolios for $\ell \le 10$. The individual results are in yellow, all the others are statistics of the yellow distribution. {\bf Right:} the histograms for $\ell=1$ and $\ell=2$.
		}\label{fig:confidence_intervals}
	\end{figure}

}

\subsubsection{Empirical Sharpe ratio results}	\label{sec:num_sharpe}

In this final numerical section we present empirical data corresponding to the theoretical Sharpe ratio results in section~\ref{sec:theortical_cont_time}.
Let us first note that the data concerns the realized Sharpe ratio, a empirical version of the theoretical Sharpe ratio in Proposition~\ref{prop:cont_time}. Also note that the theoretical result is formulated as a perturbation 
analysis but when
$\epsilon=0$ the dynamics is deterministic and has infinite Sharpe ratio; so we would like the statement to hold for values of $\epsilon$ that are large enough and test therefore realistic values of the volatility. Another question concerns how will the results carry on in a discrete setting; to this end we consider discrete markets.

\begin{figure}[htpb!]
\centering
\raisebox{2.25cm}{\parbox{3cm}{$\mu_1=\mu_2=0.15$ \\ $\sigma_1=\sigma_2=0.15$}}
\includegraphics[width=0.65\linewidth]{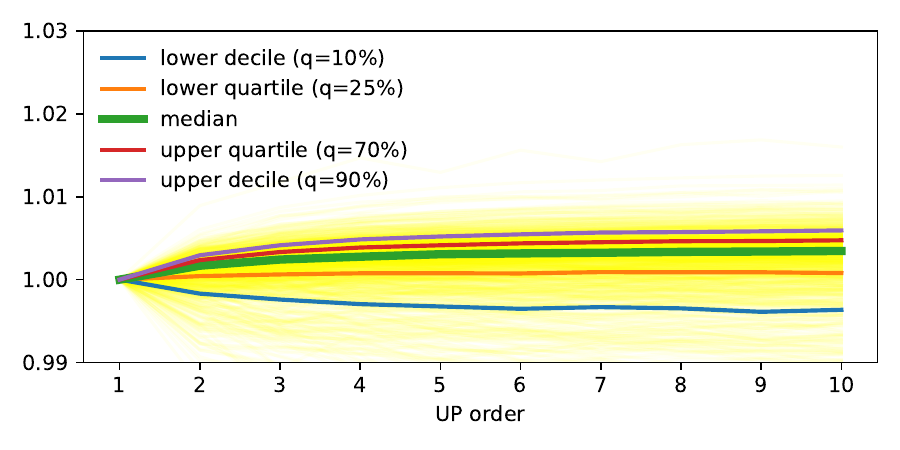}

\raisebox{2.25cm}{\parbox{3cm}{$\mu_1=\mu_2=0.3$ \\ $\sigma_1=\sigma_2=0.15$}}
\includegraphics[width=0.65\linewidth]{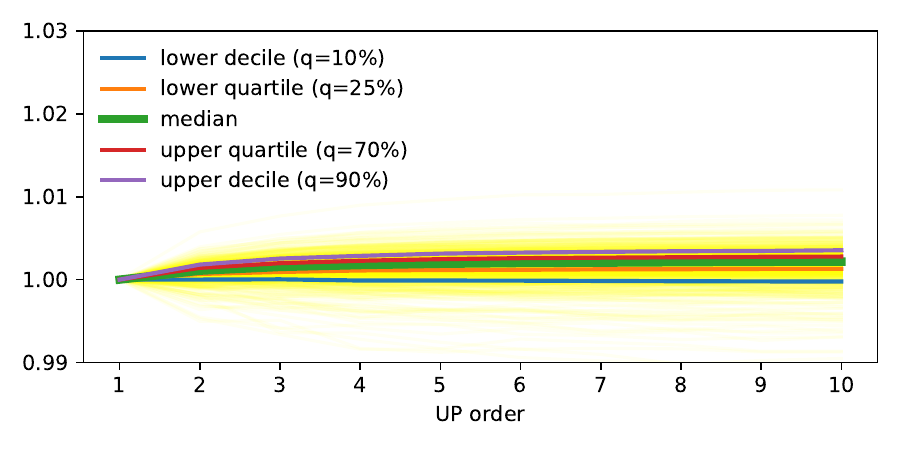}

\raisebox{2.25cm}{\parbox{3cm}{$\mu_1=\mu_2=0.15$ \\ $\sigma_1=\sigma_2=0.3$}} \ 
\includegraphics[width=0.65\linewidth]{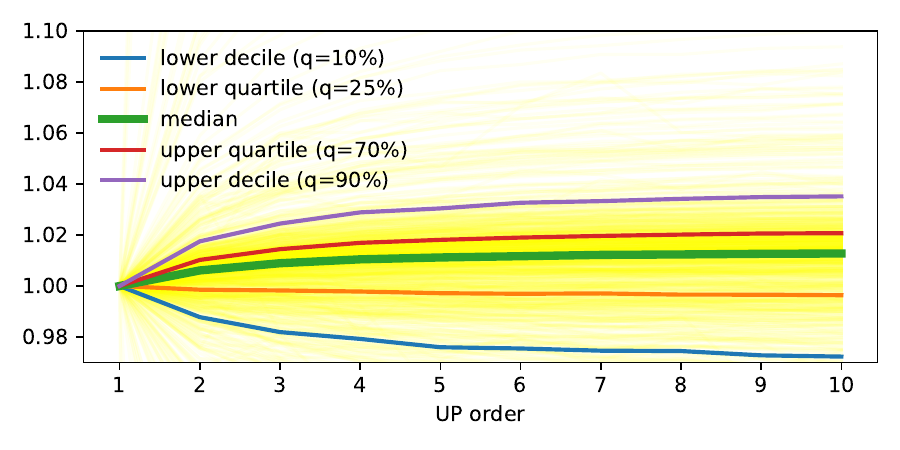}
\caption{Illustration of the section \ref{sec:num_sharpe}; we plot some quantiles of the random variable : quotient of the Sharpe ratio of $UP^\ell$ and the Sharpe ratio of  $UP^1$ for $\ell \le 10$. The values of $\mu_k$, $\sigma_k$ are given in the left of each plot.}
\label{fig:sharpe_ratios}
\end{figure}

We choose a log-normal model for $k=1,2$~:
\begin{equation}
	\frac{dS^k}{S^k}= \mu_k dt + \sigma_k dW_k(t), \ S_0^k=1, 
	\label{eq:dynamics_sharpe}
\end{equation}
where $W_k$, $k=1,2$ are two independent Brownian motions.  We take a time horizon of $10$ years divided in $T=252\times 10$ time steps; the dynamics 	\eqref{eq:dynamics_sharpe}
is sampled $1000$ times and the Sharpe ratio of all $UP^\ell$, $\ell \le 10$ is computed in each case.
We consider, as a random variable, the quotient between the Sharpe ratios of 
 $UP^\ell$ and  $UP^1$ for $\ell \le 10$ and plot some quantiles of its distribution. 
 The results  for several values of $\mu_k$ and $\sigma_k$ are given in figure~\ref{fig:sharpe_ratios}. 
 First we note that  the Sharpe ratio of $UP^\ell$ is (in median) larger than the Sharpe ratio of $UP^1$ for all choices of the parameters,  consistent with  Proposition~\ref{prop:cont_time}. We also note that when $\mu_k$ and $\sigma_k$ increase, the Sharpe ratio of $UP^\ell$ increases with respect to the Sharpe ratio of $UP^1$ so $UP^\ell$ seems more useful for assets having large volatility and average return. 

\subsubsection{Impact of the transaction costs}
\label{sec:transaction_costs}

We present below an analysis of the impact of the transaction fees 
to investigate whether HOUPs are more robust or more fragile than UPs in presence of such additional costs.

In the literature several models of transaction costs 
have been proposed 
(see for instance \cite{transaction_fees,blum1997universal} and references within)
but most of them are dependent (often linearly) on an important metric  which is the portfolio turnover, i.e. the total fraction of the portfolio traded (bought or sold).

\begin{figure}
    \centering
    \includegraphics[width=0.75\linewidth]{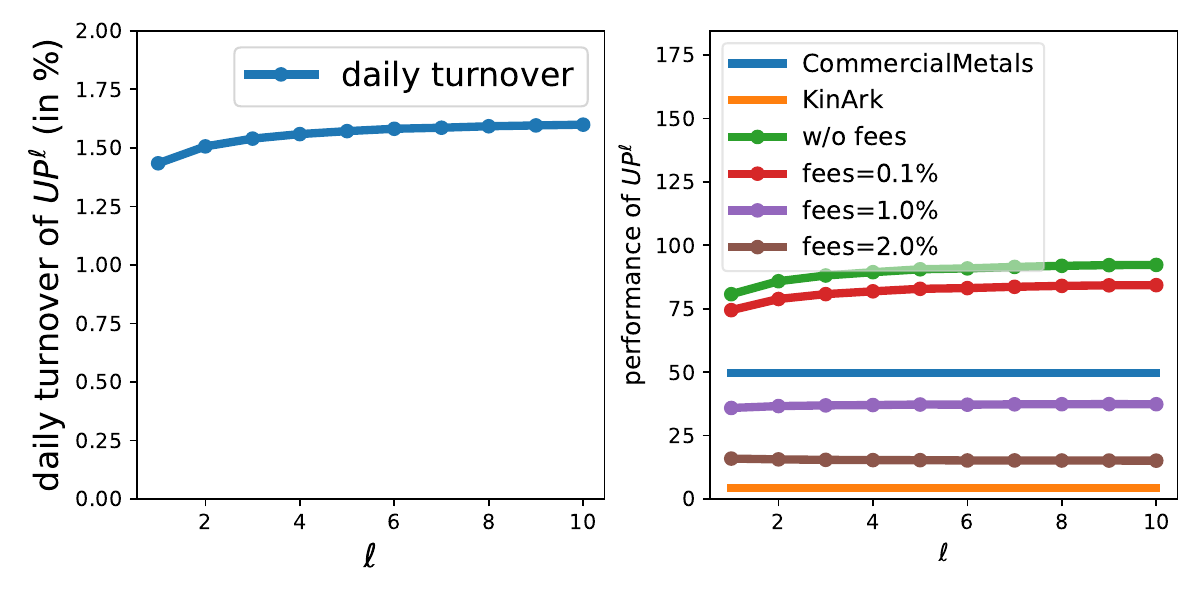}
    \includegraphics[width=0.75\linewidth]{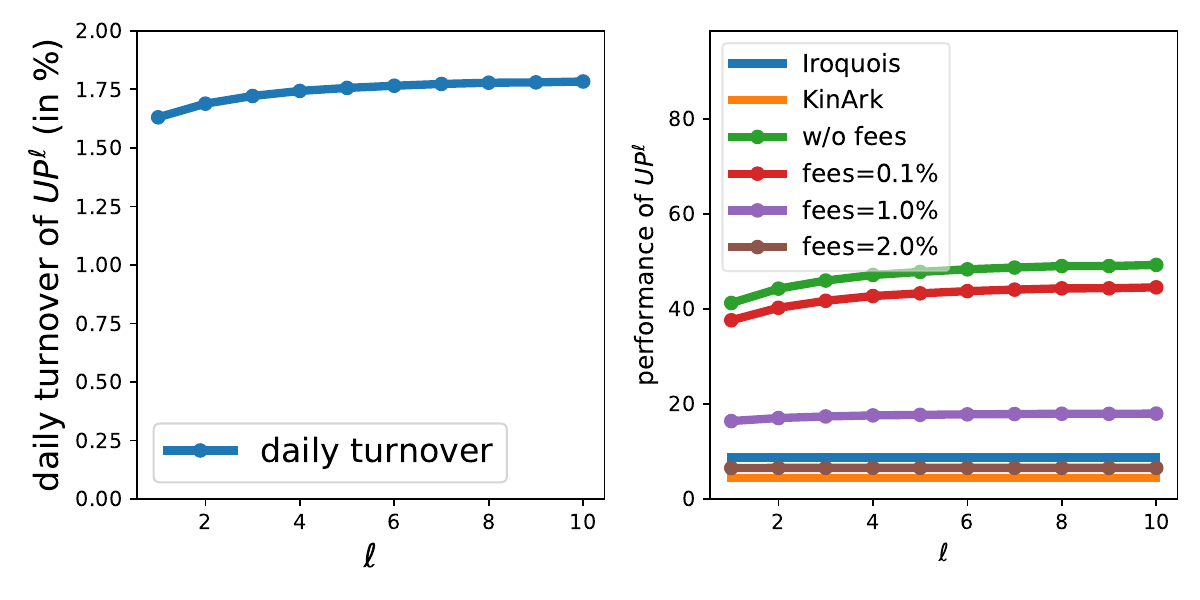}
    \includegraphics[width=0.75\linewidth]{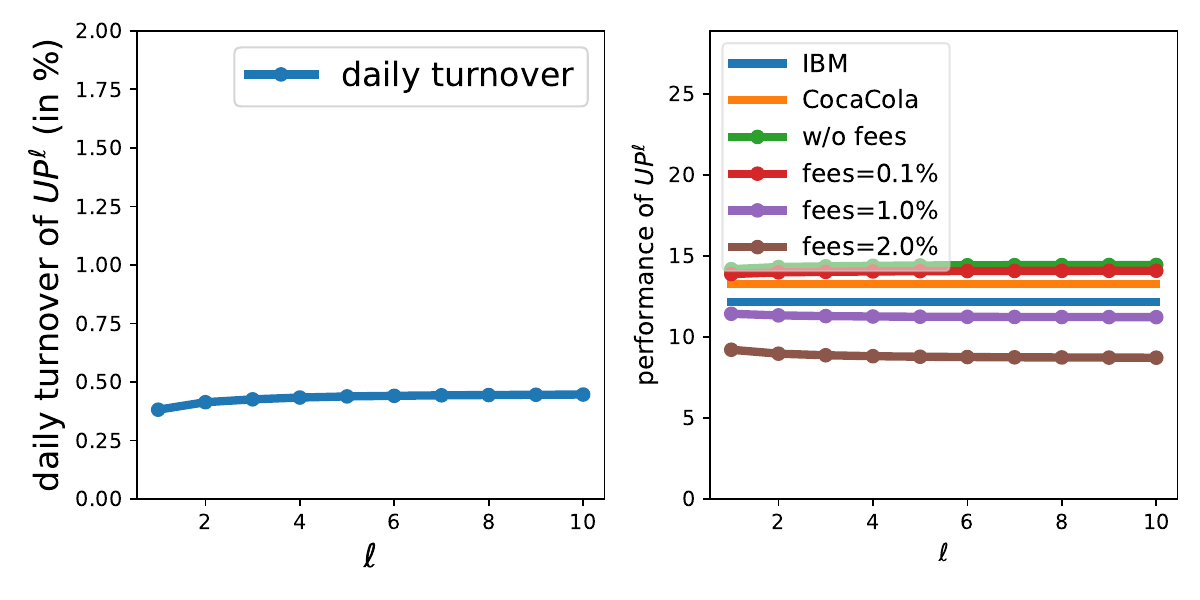}
    \includegraphics[width=0.75\linewidth]{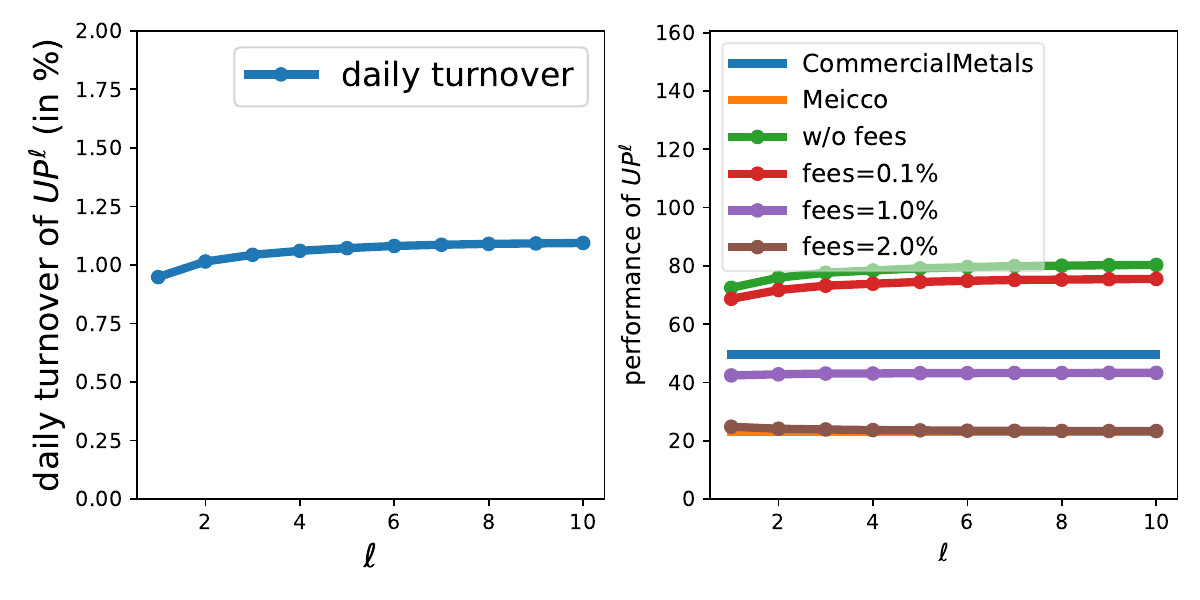}
    \caption{Impact of portfolio fees in Section~\ref{sec:transaction_costs}. Left column: the portfolio turnover, right column the portfolio value for several 
    levels of proportional fees : $0.1\%$, $1\%$ and $2\%$.}
    \label{fig:portfolio_fees}
\end{figure}

For each case in Section~\ref{sec:numerical_nyse_o}
(see also Table \ref{table:nyse_o_couples})
we calculated the turnover 
of the Universal Portfolio $UP^\ell$
with respect to the order $\ell$ and also the portfolio performance with respect to $\ell$. We consider proportional transaction costs and investigate fee levels of $0.1\%$, $1\%$ and $2\%$ assuming daily re-balancing. The level of $0.1\%$ is typical for stock trading in developed markets while the latter two are considered high values, encountered e.g., in less liquid markets. The results are presented in Figure~\ref{fig:portfolio_fees}
where we also plot for comparison, the performance of the original components of the market as horizontal lines.

We note that the mean (daily) turnover is in the range  0.25\%-1.75\% which is reasonable and increases very mildly with the portfolio order $\ell$ (left column of the plot). So the impact on the performance is also moderate as can be seen from the right column of plots and in fact we  note that for all fee levels the performance of $UP^\ell$ does not decline with respect to  that of the classical UP. However, while the fee level of $0.1\%$ still provides performances that are better than that of the best individual stock, this advantage breaks down for  fee levels above $1\%$.\footnote{As an  encouraging particular empirical result we remark that in all cases, at the fee level of $0.1\%$, the performance of $UP^{10}$ is better than the performance of $UP^1$ (classical UP) computed without any transaction cost. So the high order may compensate some of the burden of the transaction fees. Of course, there is no guarantee that this will remain true for other markets.}

So the conclusion is twofold: firstly, as expected, the transaction costs impact negatively the performance, with moderate fee levels still resulting in an advantage. Nevertheless, 
the increase in turnover with order $\ell$ is moderate and the most of the negative impact is not due to the portfolio being of high order but to the intrinsic cost of dynamic strategies with respect to static allocations.

\section{Conclusion}\label{sec:conclusion}
We introduce in this paper the high order universal portfolios (HOUP) constructed from Cover's initial suggestion by recursive additions to the underlying market.
We discuss several theoretical questions and prove that HOUP are indeed distinct from UP and can break the time invariance symmetry featured by the Cover UP. The expected optimality of the HOUP with respect to the baseline UP was investigated; we proved theoretically  that under some assumption (stated in the proposition \ref{prop:cont_time}) the Sharpe ratio increases with the order of the UP. We next performed empirical tests 
 on a dataset from the literature. In many cases implementing HOUP is more rewarding than the UP; of course, these empirical results are not always guaranteed and situations can occur when HOUPs perform worse than the UP; nevertheless this first joint theoretical and numerical evidence appears positive. Further studies could shed additional light onto the performances of the high order universal portfolios introduced here and their applicability domain.
%

\subsection{Ethical Statement}
The author does not declare any conflicts of interests. The research did not involve and human participants and/or animals. 

\appendix

\section{Python code for the proof of Proposition \ref{prop:houp_not_all_equal}
} \label{appendix:python_code_for_proof}
We list below the short Python code that was used in the proof of Proposition~\ref{prop:houp_not_all_equal}.
\lstset{texcl=true,basicstyle=\small\sf,commentstyle=\small\rm,mathescape=true,escapeinside={(*}{*)}}
\begin{lstlisting}
	import sympy as sym
	a1, a2, b1,b2,c1,c2 = sym.symbols('a1 a2 b1 b2 c1 c2')
	c1=a1/2+b1/2
	c2 = (2*a1*a2+2*b1*b2+a1*b2+a2*b1)/3/(a1+b1)
	up1_2=(2*a1*a2+2*b1*b2+a1*b2+a2*b1)/6
	up1_2.simplify()
	#a1*a2/3 + a1*b2/6 + a2*b1/6 + b1*b2/3
	up2_2=((a1+b1+c1)*(a2+b2+c2)+a1*a2+b1*b2+c1*c2)/12
	up2_2.simplify()
	#23*a1*a2/72 + 13*a1*b2/72 + 13*a2*b1/72 + 23*b1*b2/72
\end{lstlisting}

\section{Self-contained proof of Lemma~\ref{lemma:simplex}}
\label{appendix:dirichlet}

By symmetry all $\int_{\Scal_K} w_i  dw$ are equal for $i=1,...,K$. Since their sum is  $\int_{\Scal_K} \sum_{k=1}^K w_k  dw =\int_{\Scal_K} 1 dw = 1$, the first identity follows.

For the second identity denote $\beta_K =  \int_{\Scal_K} (w_1)^2dw$. Of course, for all $j$ we have $\int_{\Scal_K} (w_j)^2dw= \beta_K$.
Let us now compute $ \beta_K$. We will sample $\Scal_K$ as follows: we sample at random $x_1, ..., x_{K-1}$ uniform from $[0,1]$, add to their set the values $0$ and $1$ and order the set to obtain $y_1=0 \le y_2 \le ... \le y_K \le y_{K+1}=1$. Then $y_{k+1}-y_k$ is a uniform sampling from $\Scal_K$. In particular $w_1 = \min\{x_1, ..., x_{K-1}\}$. The minimum can be any of the $x_k$, so by symmetry~: 
$\int_{\Scal_K} (w_1)^2 dw = (K-1) \int_0^1  (x_1)^2 \int_{x_1}^1  ... \int_{x_1}^1 dx_{K-1} dx_2 dx_1 = (K-1) \int_0^1  (x_1)^2 (1-x_1)^{K-2} dx_1 = \frac{2}{K(K+1)}$.
This means that $\beta_K= \frac{2}{K(K+1)}$.

On the other hand, by symmetry~:
$\frac{1}{K} = \int_{\Scal_K} w_1  dw = \int_{\Scal_K} w_1 \sum_{k=1}^K w_k dw 
= \beta_K + (K-1)\alpha_K$ and the value of $\alpha_K$ follows together with the last  identity.

\end{document}